\documentclass[a4paper, amsfonts, amssymb, amsmath, reprint, showkeys, nofootinbib, twoside, superscriptaddress]{revtex4-1}
\usepackage[english]{babel}
\usepackage[utf8]{inputenc}
\usepackage[colorinlistoftodos, color=green!40, prependcaption]{todonotes}
\usepackage[pdftex, pdftitle={Article}, pdfauthor={Author}]{hyperref} 
\usepackage[normalem]{ulem}
\usepackage{color}
\usepackage{subfigure}
\usepackage{enumitem}
\usepackage{amsmath}
\DeclareMathOperator\arctanh{arctanh}

\newcommand{\LCDM}{$\Lambda$CDM}
\def\bel#1{\begin{equation} \label{#1}}

\def\mpl{M_{\rm pl}}

\def\be{\begin{equation}}
\def\ee{\end{equation}}
\def\bea{\begin{eqnarray}}
\def\eea{\end{eqnarray}}

\def\ltap{\ \raise.3ex\hbox{$<$\kern-.75em\lower1ex\hbox{$\sim$}}\ }
\def\gtap{\ \raise.3ex\hbox{$>$\kern-.75em\lower1ex\hbox{$\sim$}}\ }
\def\gl{\ \raise.5ex\hbox{$>$}\kern-.8em\lower.5ex\hbox{$<$}\ }
\def\roughly#1{\raise.3ex\hbox{$#1$\kern-.75em\lower1ex\hbox{$\sim$}}}

\def\mpl{M_{\rm pl}}

\newcommand{\comments}[1]{}

\newcommand{\ben}{\begin{enumerate}}
\newcommand{\een}{\end{enumerate}}
\newcommand{\bi}{\begin{itemize}}
\newcommand{\ei}{\end{itemize}}
\newcommand{\ba}{\begin{align}} 
\newcommand{\ea}{\end{align}}

\def\beq{\begin{equation}}
\def\eeq{\end{equation}}
\def\bea{\begin{eqnarray}}
\def\eea{\end{eqnarray}}

\begin{document}
\title{A Faster Growth of Perturbations in an Early Matter Dominated Epoch: \\  Primordial Black Holes  and Gravitational Waves}

\author{Subinoy Das}
    \email[]{subinoy@iiap.res.in}
    \affiliation{Indian Institute of Astrophysics, Bengaluru, Karnataka 560034, India.}
 \author{Anshuman Maharana}
    \email[]{anshumanmaharana@hri.res.in}
    \affiliation{Harish-Chandra Research Institute, HBNI, Allahabad, Uttar Pradesh 211019, India.}
  \author{Francesco Muia}
    \email[]{fm538@cam.ac.uk}
    \affiliation{DAMTP,  Wilberforce Road, Cambridge CB3 0WA, United Kingdom.}   


\begin{abstract}

We present a scenario for fast growth of cosmological perturbations;  $\delta(t) \sim a(t)^s$, $a(t)$ being the scale factor, with $s > 10$ for the numerical examples reported in this article. The basic ingredients of the scenario are an early matter dominated era and the dark fermion which experiences a scalar mediated force during the epoch. Both of these arise in string/supergravity models. The fast growth occurs for sub-horizon density perturbations of the dark fermion. The fast growth has a rich set of phenomenological implications. We outline implications for the formation of primordial black holes and the production of gravitational waves. Primordial black holes in the sub-lunar mass range (which are ideal dark matter candidates) can be produced. Gravitational waves can be produced in a wide range of frequencies due to second order scalar perturbations and due to evaporation and merger of primordial black holes.

\end{abstract}


\maketitle

\tableofcontents

\setcounter{footnote}{0}

\section{Introduction}

\label{sec:intro}
Understanding inhomogeneities in the observed universe is one of the central challenges in cosmology. The tendency for gravitational collapse is much weaker in an expanding universe than in flat space. Therefore, mechanisms for the fast growth of cosmological perturbations are of much interest. For instance, in matter domination an overdensity $\delta = \delta\rho/\rho$ (where $\rho$ is the energy density and $\delta\rho$ is the perturbed energy density) grows linearly with the scale factor: $\delta(t) \propto a(t)$. Here, we will present a scenario for much faster growth.

This growth will take place in a pressure-less fluid composed of non-relativistic fermionic particles $\psi$ in the early universe, prior to \textit{Big Bang Nucleosynthesis} (BBN).\footnote{For a recent review of pre-BBN physics see~\cite{Allahverdi:2020bys}.} The scenario has two essential ingredients: \textit{i)} an \textit{Early Matter Domination} (EMD) era;\footnote{Usually an epoch of EMD leads to linear growth of overdensities, similar to what happens in the dark matter dominated epoch in the late universe~\cite{Georg:2016yxa, Georg:2017mqk, Erickcek:2011us, Redmond:2018xty}.} \textit{ii)} an interaction between the $\psi$ particles mediated by a massive scalar field $\phi$ in the EMD epoch. These ingredients arise naturally in string/supergravity models.

We present explicit examples of background solutions incorporating the interactions and then study perturbations to exhibit the rapid growth. The physical reason for the fast growth is the scalar field mediated force, as $\psi$ particles attract each other with a stronger than gravitational interaction. We will also initiate a study of the phenomenological implications of this scenario. These will include the production of \textit{Primordial Black Holes} (PBHs) and \textit{Gravitational Waves} (GWs).\\

Let us start by discussing the generic features of the essential ingredients of our setup.\\
 
\noindent {\it{Early Matter Domination}:}

A generic feature of string/supergravity models is the existence of moduli, i.e. gravitationally coupled scalar fields whose \textit{vevs} parametrise the size and shape of the extra-dimensions. These are massless at tree level and typically acquire masses due to higher-order corrections or loop effects. In many cases, moduli masses are set by the scale of supersymmetry breaking and it is well below the Hubble scale at the end of inflation. This implies a displacement of the scalar fields from their late time minima. The displaced scalars oscillate about their minima when the Hubble constant falls below their masses, leading to epochs of matter domination (see e.g.~\cite{Coughlan:1983ci, hep-ph/9308292, hep-ph/9308325, hep-ph/9507453, Cicoli:2016olq,  1906.03025, Erickcek:2011us}  and~\cite{1502.07746} for a review). 

To accommodate the successful predictions of BBN, an EMD epoch has to end with the universe reheating above $\sim 3 \, \rm{MeV}$. This happens with the decay of the constituents of the universe during the EMD epoch. We note that while an oscillating modulus is one of the most natural ways to enter an epoch of EMD, this is not necessary for our scenario, which is insensitive to the exact nature of the particle causing the EMD epoch (we will denote quantities associated with this by a sub/superscript ‘b', for \textit{background}).\\

\noindent {\it{Scalar Field Mediated Interactions}:}

The other key element is a hidden sector fermion ($\psi$) experiencing a scalar ($\phi$) mediated force. Cosmological effects of a hidden fermion experiencing a scalar mediated force have been extensively studied in recent years, see e.g.~\cite{Das:2005yj, Amendola:1999er, Vagnozzi:2021quy, Tsai:2021irw, Savastano:2019zpr,  Amendola:2017xhl, Damour:1990tw, hep-th/9408025, Amendola:1999er, gr-qc/0108016, astro-ph/0303145,
astro-ph/0208032, astro-ph/0306343, astro-ph/0212518, astro-ph/0307350, Amendola:2003wa, Domenech:2021uyx}.

We will take $\psi$ to be part of a hidden sector. Its interactions with the visible sector will be feeble\footnote{Therefore the usual fifth-force bounds are not relevant.}, but it will be strongly interacting with $\phi$. In the Einstein frame, the scalar couples to the trace of the energy-momentum of $\psi$: $g^{\mu \nu} T_{\mu \nu}^{(\psi)} = \rho_{\psi} - 3 p_\psi$, where $\rho_\psi$ and $p_\psi$ are the energy density and the pressure of the $\psi$ fluid respectively. At the level of cosmological fluids, this implies the non-conservation of stress tensors of individual components. The violation is proportional to the product between the trace of the energy-momentum tensor of the $\psi$ component and the gradient of the mediating scalar. At the background level, once the scalar field is oscillating about its minimum (i.e. when $H \lesssim m_\phi$, where $H$ is the Hubble parameter) the fluids reach a scaling regime in which each component redshifts as matter and the energy exchange between fluids stops. However, the presence of the coupling leads to an attractive force between $\psi$ particles. This force leads to the fast growth of $\psi$ perturbations.\footnote{Note that in~\cite{Savastano:2019zpr} a similar mechanism causes matter perturbations to grow much faster compared to what happens in \LCDM, i.e. $\delta(t) \propto a(t)^{1.62}$, even in a radiation dominated epoch. The present work is much inspired by this.}\\
 
We note that, when the coupling is non-vanishing, the background dynamics of the scalar field $\phi$ is affected. A complete study of the system would then require tracking the dynamics starting from the end of inflation, or fine-tuning the initial conditions. This is an interesting and important issue in all models that feature similar couplings~\cite{Das:2005yj, Amendola:1999er, Vagnozzi:2021quy, Tsai:2021irw, Savastano:2019zpr,  Amendola:2017xhl, Damour:1990tw, hep-th/9408025, Amendola:1999er, gr-qc/0108016, astro-ph/0303145,
astro-ph/0208032, astro-ph/0306343, astro-ph/0212518, astro-ph/0307350, Amendola:2003wa, Domenech:2021uyx}. Alternatively, one can consider scenarios in which a change in the equation of state $w_\psi$ of the $\psi$ component takes place at the energy scale of interest, going from $w_\psi = 1/3$ to $w_\psi = 0$. Given the form of the coupling, $\propto \rho_\psi (1 - 3 w_\psi)$, this would imply that the coupling is turned on only when the equation of state deviates from $1/3$. There are in principle various possible ways to achieve this, such as scenarios where the $\psi$ particles get mass from a hidden sector symmetry breaking when a hidden scalar field acquires a vacuum expectation value~\cite{Gehrlein:2019iwl, Shelton:2010ta}. Before the symmetry breaking the $\psi$ particles would essentially be massless and behave as relativistic degrees of freedom. In general the time scale of such a symmetry breaking is much smaller than Hubble time, so one would expect a rapid transition in the $\psi$ equation of state as assumed in this paper. The second approach is a time dependent coupling constant which  is a function of the scalar vacuum expectation value~\cite{Hinterbichler:2010es}. In this well studied symmetron-based model, one naturally turns on the fifth-force or a non-zero coupling between the fermion and the scalar  when the universe starts expanding followed by a symmetry breaking in a hidden scalar sector. It is instructive to note that the actual model building of our scenario is not the focus of this work. In this paper, we will use a phenomenological approach, parametrising the equation of state to keep track of the coupling. We leave a study of the microscopic realisation of such a transition to a future work.\\

Before closing the introduction, let us briefly mention the possible phenomenological implications of our scenario. Estimates of the scales involved give production of PBHs
in the sub-lunar mass range. These are known to be ideal candidates to constitute all of dark matter. GWs will be produced with the scalar field acting as a source for the perturbations and from the dynamics of the PBHs produced. This leads to GWs in a wide range of frequencies (from $10^{-3}$ to $10^{15}$ Hz).\\
 
This paper is structured as follows. In Sec.~\ref{background}, we describe our setup and obtain background solutions.
In Sec.~\ref{sec:Perturbations} we analyse perturbations and exhibit their fast growth. We outline phenomenological implications
in Sec.~\ref{pheno}, leaving detailed explorations for future work. We conclude in Sec.~\ref{conclude}.

\section{Coupled  Dynamics in the Early Universe}
\label{background}

In this section, we first describe the equations that govern the dynamics of our system. We then present the background (homogeneous) cosmology in which perturbations will exhibit fast growth.  The solution settles into a matter dominated phase within a few e-foldings of cosmological expansion: it is in this epoch that the fast growth of perturbations takes place.

As described in the introduction, the early universe we consider will have three constituents: a background component redshifting as matter ($b$)\footnote{E.g. an oscillating modulus.}, a dark sector fermion ($\psi$) and the scalar force mediator ($\phi$). The first two will be described by cosmological fluids, the latter by its equation of motion. In general, a coupling to the scalar via the trace of the stress tensor can exist for both `b' and `$\psi$'. The system is described by the following equations (see~\cite{Amendola:2003wa} and references therein):
\begin{align}
\label{eq:GeneralConservationBackground}
 \nabla^{\mu} T_{\mu \nu}^{b} &= - {\beta_{b}(\phi) \over M_{\rm pl} } g^{\rho \sigma} T_{\rho \sigma}^{b} \nabla_{\nu} \phi \,, \\
 \label{eq:GeneralConservationPsi}
 \nabla^{\mu} T_{\mu \nu}^{\psi} &= - {\beta_{\psi}(\phi) \over M_{\rm pl}}g^{\rho \sigma} T_{\rho \sigma}^{\psi} \nabla_{\nu} \phi \,, \\
 \label{eq:GeneralKGEquation}
 \left( \square  + m^{2} \right) \phi &=   {\beta_{b}(\phi) \over M_{\rm pl}} g^{\rho \sigma} T_{\rho \sigma}^{b} + {\beta_{\psi}(\phi) \over M_{\rm pl}} g^{\rho \sigma} T_{\rho \sigma}^{\psi} \ ,
\end{align}
 and
 \begin{equation}
 \label{eq:Einstein}
  M_{\rm pl}^{2}  G_{\mu \nu} =  T_{\mu \nu},
\end{equation}
where $g_{\mu \nu}$ is the spacetime metric, $G_{\mu \nu}$ is the Einstein tensor and $T_{\mu \nu} = T_{\mu \nu}^{(b)} + T_{\mu \nu}^{(\psi)} + T^{(\phi)}_{\mu \nu}$ is the total stress-energy tensor.

The quantities $\beta_{b}(\phi)$ and $\beta_{\psi}(\phi)$ are  (field dependent) coupling constants and $M_{\rm pl}$ is the reduced Planck mass. Note that large values of the coupling constants $(\beta_{i}(\phi) \gg 1)$ imply that the scalar mediates a force that is stronger than gravity. Hierarchies in the strengths of the couplings can arise naturally in string models as a result of physical separation in the extra-dimensions between different sectors (see e.g.~\cite{Acharya:2018deu} and references therein for a recent discussion in the context of quintessence models). The goal of this paper is to exhibit the phenomenon of fast growth in a specific setting and thereby provide proof of the concept. Hence we will consider a constant coupling $\beta_\psi(\phi) \equiv \beta_\psi \gg 1$, while we take $\beta_b(\phi) = 0$ (a detailed exploration of the dynamics treating both the couplings as parameters is left for future work).

The stress tensors $T_{\mu \nu}^{(b)}$ and $T_{\mu \nu}^{(\psi)}$ will be taken to be of the perfect fluid form. The component `b' seeds the matter dominated epoch; we will take $w_b =0$. As mentioned in the introduction, the $\psi$ component will make a transition from being relativistic (at early times) to becoming non-relativistic within a few e-foldings of the expansion of the universe from the start of our numerical evolution. This transition sets the form of its (time-dependent) equation of state\footnote{We will describe its precise form soon.}, $w_{\psi}$. The scalar stress tensor $T^{(\phi)}_{\mu \nu}$ is given by $T^{(\phi)}_{\mu \nu} = \nabla_\mu \phi \nabla_\nu \phi - g_{\mu \nu} \left(\frac{1}{2} g^{\lambda \rho} \nabla_\lambda \phi \nabla_\rho \phi - V(\phi)\right)$. For simplicity, we work with $V(\phi) = {1 \over 2} m_\phi^{2} \phi^{2}$.


Finally, the matter dominated epoch has to end before BBN. The time of the decays is controlled by the widths of the fields. We will treat the widths $\Gamma_{b}$, $\Gamma_{\psi}$ and $\Gamma_\phi$ as phenomenological parameters in our study.

\subsection{Background Dynamics}
\label{sec:BackgroundDynamics}

Next, let us turn to the background solution. Using the number of e-foldings $N = \int H dt$ as the evolution variable, for a homogeneous background Eq.s~\eqref{eq:GeneralConservationBackground}-\eqref{eq:GeneralKGEquation}  become
\begin{align}
\label{eq:ConservationBackground}
&\rho_b' + 3 H \rho_b = 0 \,, \\
\label{eq:ConservationPsi}
&\rho_\psi' + 3 (1 + w_\psi) H \rho_\psi = - \beta_\psi (1 - 3 w_\psi) \rho_\psi \phi' \,, \\
\label{eq:KGEquation}
&H H' \phi' + H^2 \phi'' + m_\phi^2 \phi + 3 H^2 \phi' = \beta_\psi (1 - 3 w_\psi) \rho_\psi \,,
\end{align}
where we have used that $p_b = 0$, $p_\psi = w_\psi \rho_\psi$, and the primes denote derivatives with respect to $N$. The  Friedmann equation reads:
\begin{equation}
H^2 = \frac{2 \left(\rho_b + \rho_\psi + \frac{1}{2} m_\phi^2 \phi^2 \right)}{6 - \phi'^2} \,.
\label{eq:FriedmannEquation}
\end{equation}
An explicit form of the background solution will be presented for two benchmark values of $\beta_{\psi} =20, 30$. We will refer to these as examples 1 and 2 respectively.

\begin{figure}[h!]
  \includegraphics[width=0.5\textwidth]{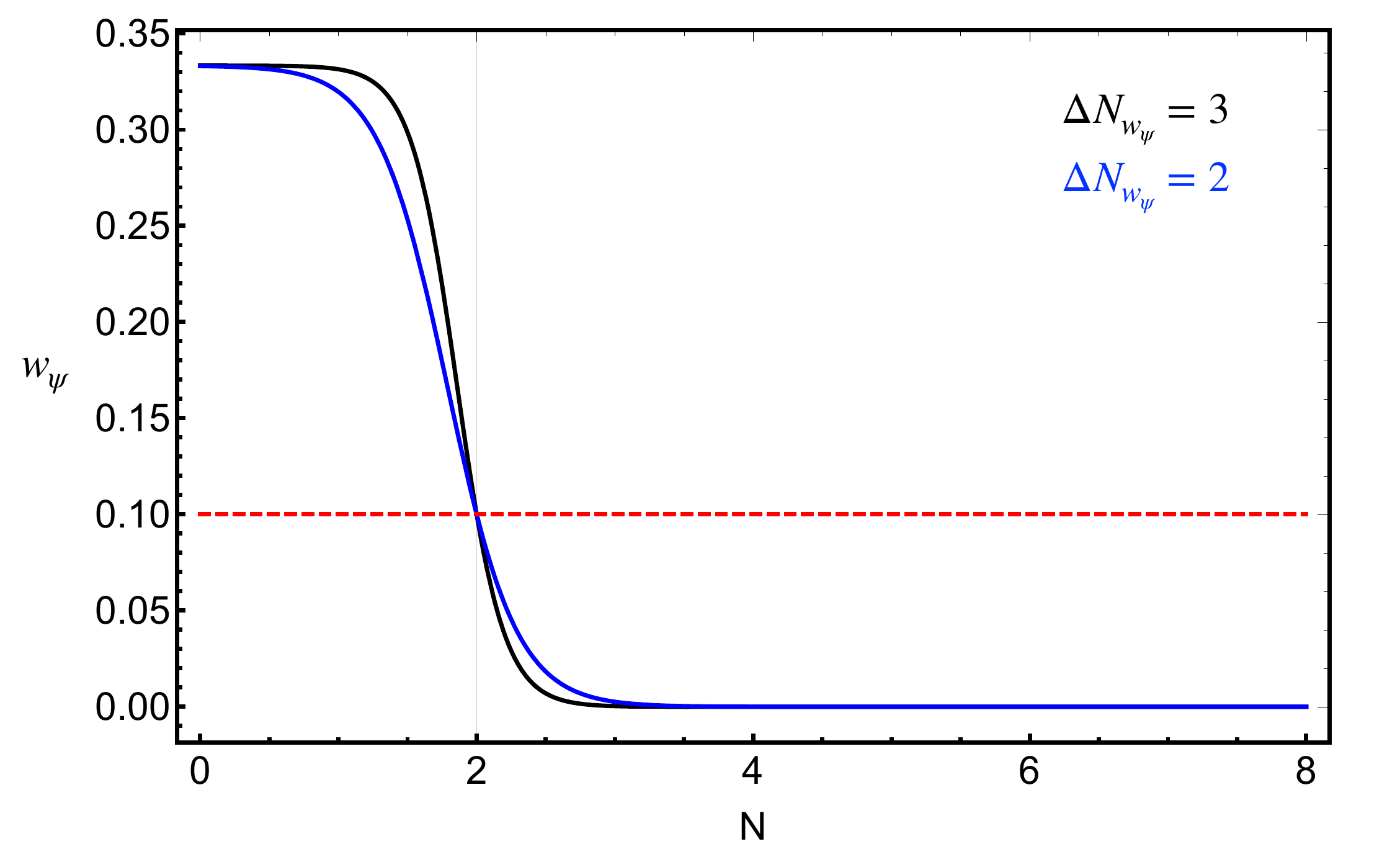}
\caption{Equation of state for $N_{\rm NR} = 2$ and $\Delta N_{w_\psi} = 2,3$.}
\label{fig:EoS}
\end{figure}

As already mentioned, we take a phenomenological approach to describe the equation of state of the $\psi$ component. Therefore, we parametrise $w_\psi$ in terms of the e-folding $N_{\rm NR}$ at which $w_\psi = 0.1$ and the width of the transition $\Delta N_{w_\psi}$:
\begin{equation}
\label{eq:EoS}
w_\psi = \frac{1}{6} \left(- \tanh\left(\Delta N_{w_\psi} (N - (N_{\rm NR} - \delta N))\right)\right) \,,
\end{equation}
where $\delta N = \arctanh(0.6/\Delta N_{w_\psi})$ is adjusted so that $w_\psi = 0.1$ at $N_{\rm NR}$. In the examples reported below, we fix $N_{\rm NR} = 2$ and use two values for $\Delta N_{w_\psi}$: $\Delta N_{w_\psi} = 3$ (example 1) and $\Delta N_{w_\psi} = 2$ (example 2), see Fig.~\ref{fig:EoS}.

At early times, the Hubble constant is much greater than the mass of the scalar and $w_{\psi} =1/3$. The former implies that the friction term in the left hand side of Eq.~\eqref{eq:KGEquation} vanishes and the latter implies that the right hand side of the same equation vanishes. This implies that the scalar is at rest at $\phi = \phi_{\rm in}$ at early times: it contributes to the energy density of the universe as a result of its initial misalignment. This energy density is $\rho_{\phi, \rm{in}} = {1 \over 2} m_\phi^{2} \phi_{\rm in}^2$. We will track the evolution starting from the point when the ‘initial’ Hubble parameter is $H_{\rm in} \sim m_\phi$ (this will be taken to correspond to $N=N_{\rm in}=0$). The other initial conditions that need to be specified are the initial energy densities in $b$ and $\psi$, $\rho_{\psi, \rm{in}}$ and $\rho_{b, \rm{in}}$. In the explicit examples that we will report, we use $\phi_{\rm{in}} = 0.1 \mpl$ and $\rho_{\psi, \rm{in}}/\rho_{b, \rm{in}} = 1$, $m_\phi/H_{\rm in} \simeq 1.22$. The exact value of $m_\phi/H_{\rm in}$ does not affect the results reported below, as long as the field is initially at rest.

\begin{figure}[h!]
  \includegraphics[width=0.5\textwidth]{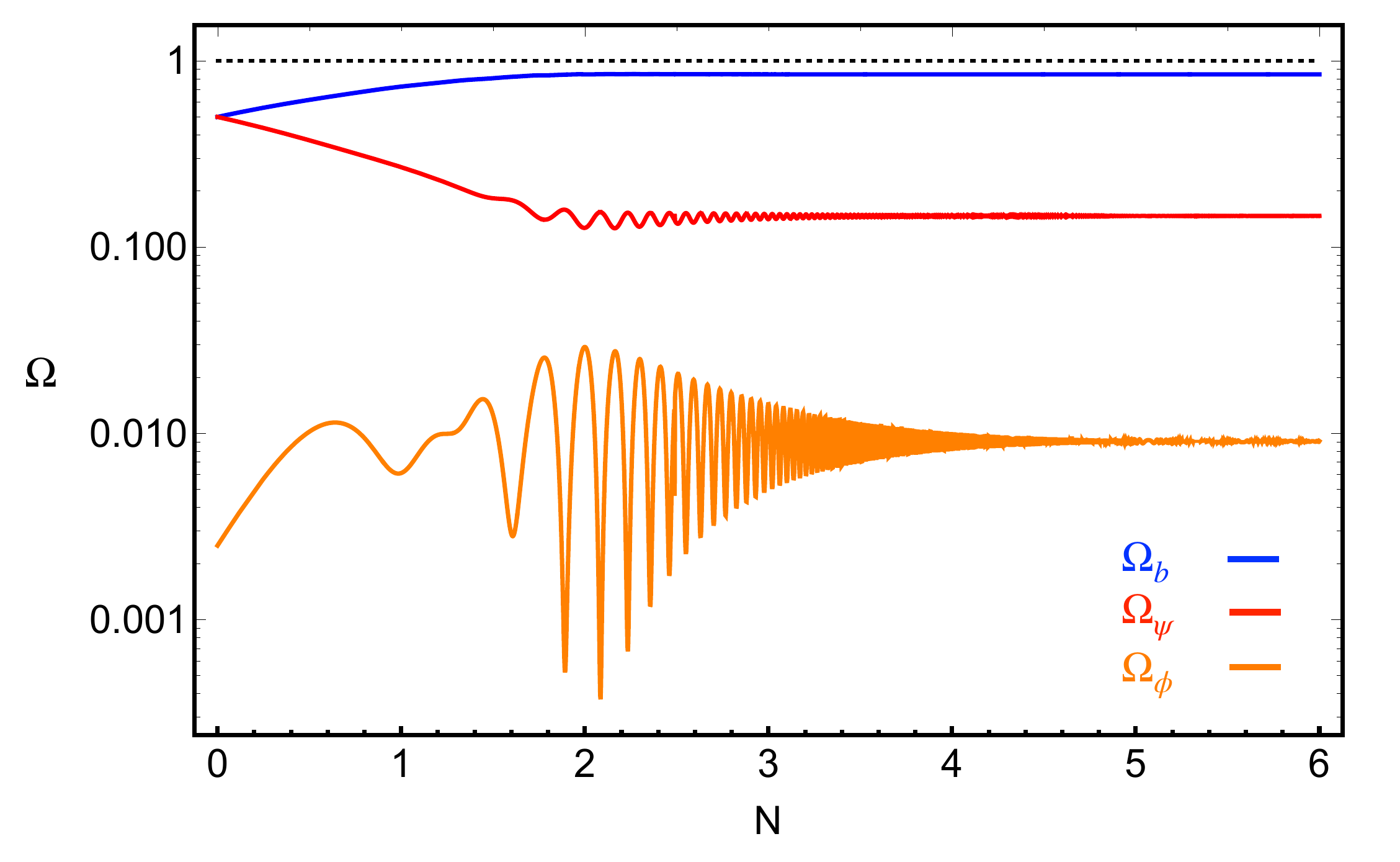}
\caption{Scaling regime for example 1, with $\beta_\psi = 20$ and $\Delta N_{w_\psi} = 3$.}
\label{fig:ScalingRegime1}
\end{figure}

\begin{figure}[h!]
  \includegraphics[width=0.5\textwidth]{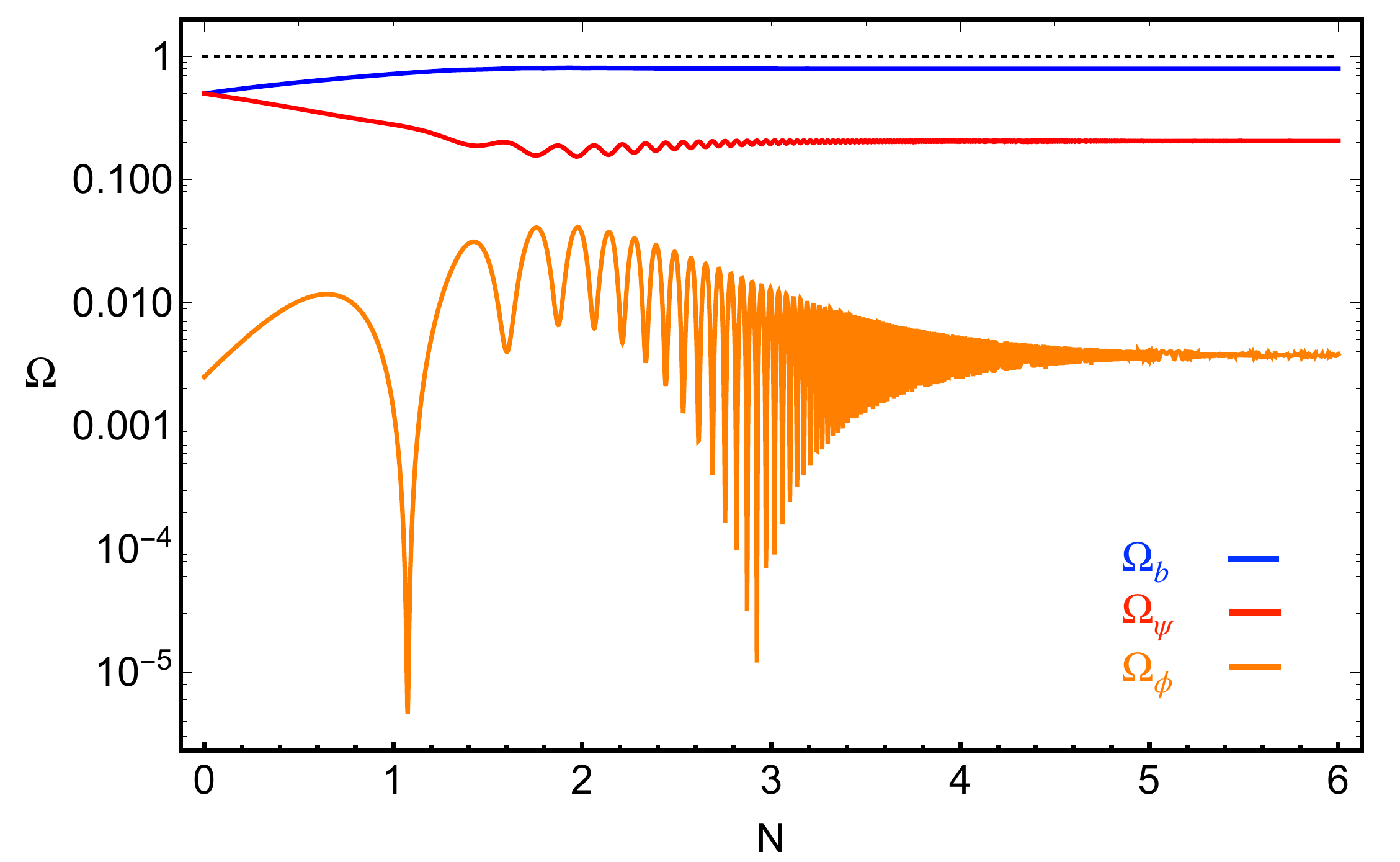}
\caption{Scaling regime for example 1, with $\beta_\psi = 30$ and $\Delta N_{w_\psi} = 2$.}
\label{fig:ScalingRegime2}
\end{figure}

\begin{figure}[h!]
  \includegraphics[width=0.45\textwidth]{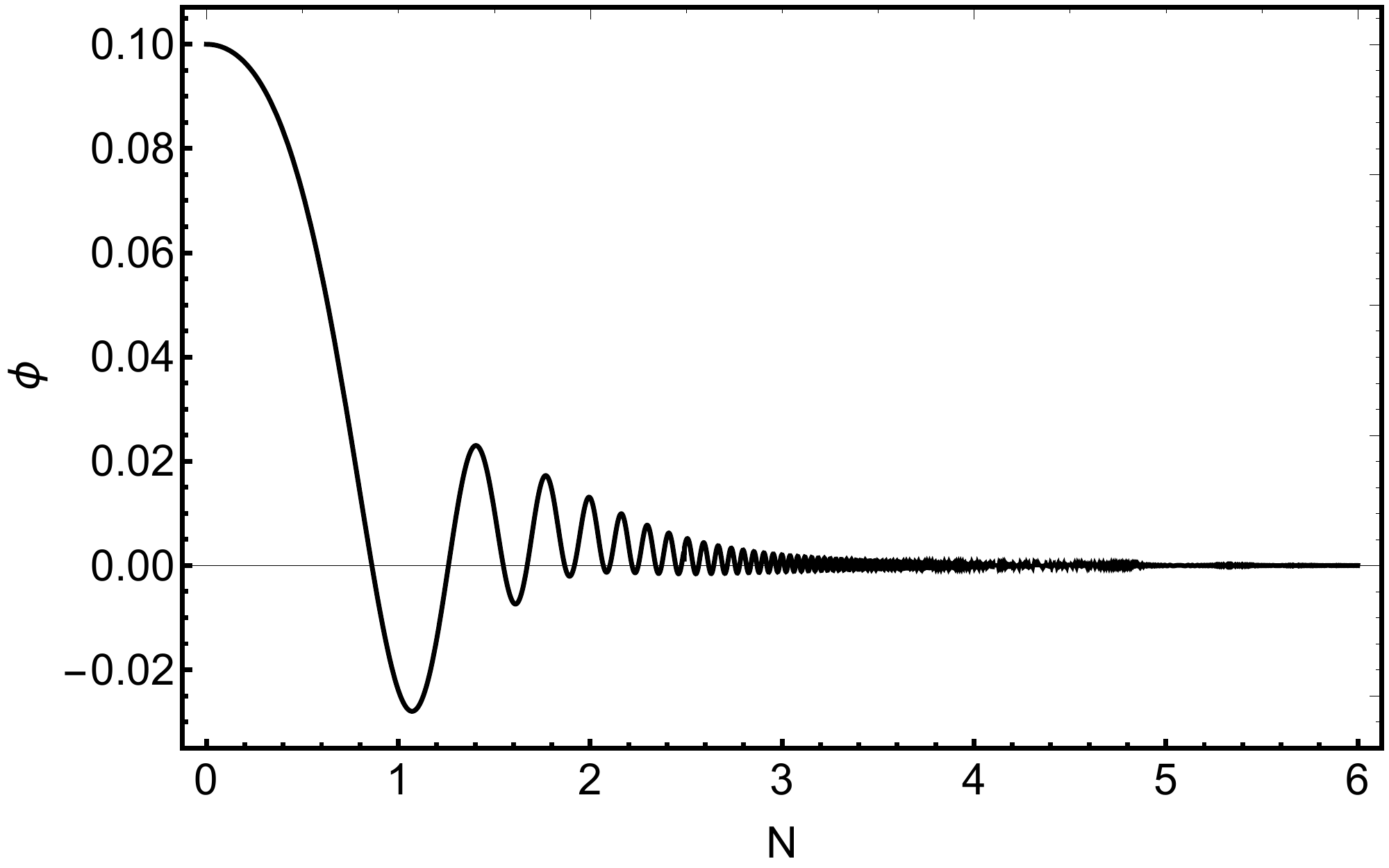}
\caption{Scalar field dynamics for example 2, with $\beta_\psi = 20$ and $\Delta N_{w_\psi} = 3$.}
\label{fig:FieldDynamics}
\end{figure}

As it is easy to guess (given that all components behave as matter when interactions are switched off), the system quickly settles into a scaling regime in which the energy densities of all the components redshift as $1/a^3$. We plot the fractional energy densities 
\begin{equation}
\Omega_b = \frac{\rho_b}{\rho} \,, \quad \Omega_{\psi} = \frac{\rho_\psi}{\rho} \,, \quad \Omega_\phi = \frac{\rho_\phi}{\rho} \,,
\end{equation}
and $\rho = \rho_b + \rho_\psi + \rho_\phi$, for our benchmark examples in Fig.~\ref{fig:ScalingRegime1} and Fig.~\ref{fig:ScalingRegime2}. The evolution of the scalar field for example 1 $(\beta_{\psi} =20, \Delta N_{w_\psi} = 3)$ is shown in Fig.~\ref{fig:FieldDynamics} (the $\phi$ evolution is very similar for example 2).


\section{Fast Growth of Perturbations}
\label{sec:Perturbations}

Having obtained the homogeneous background in the previous section, we now turn to the study of perturbations in the background. We will see that there is a fast growth of the perturbations in the matter dominated epoch which the background solutions asymptote to.

\begin{figure}[h!]
  \includegraphics[width=0.45\textwidth]{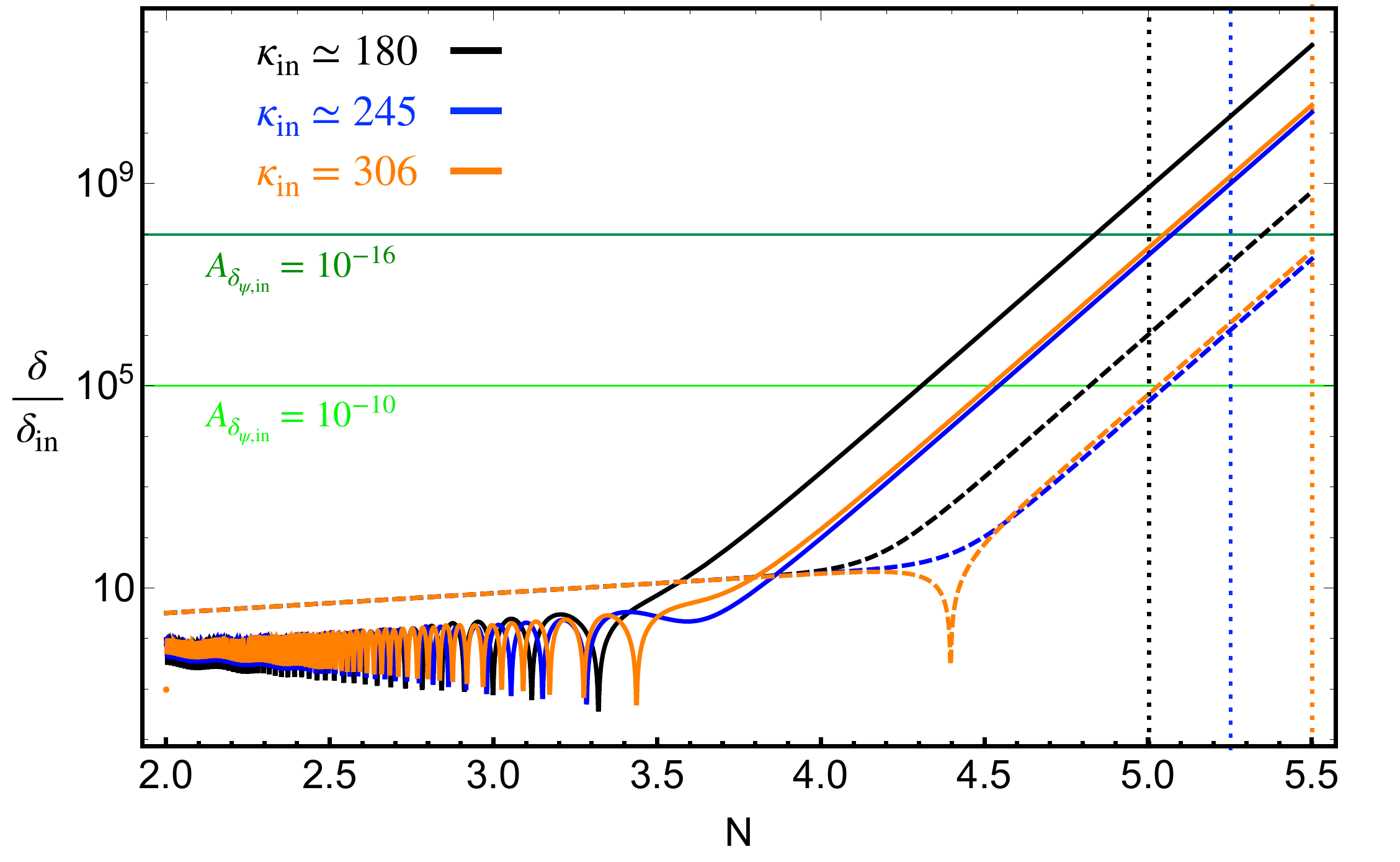}
\caption{Perturbations $\delta_\psi$ and $\delta_b$ for example 1. Solid lines represent $\delta_\psi$ and dashed lines represent $\delta_b$ for the various modes. The dotted lines indicate the maximum $N$ at which the solution with the corresponding color can be trusted, as Eq.~\eqref{eq:PerturbationDeltaPsi} and Eq.~\eqref{eq:PerturbationDeltaB} require $k/a \gtrsim m_\phi$. All the modes get to $\delta/\delta_{\rm in} \gtrsim 10^9$ in the regime of validity of the equations. When perturbations hit the light (for $A_{\delta_{\psi, \rm in}} = 10^{-10}$) and dark (for $A_{\delta_{\psi, \rm in}} = 10^{-16}$) green horizontal lines they go non-linear.}
\label{fig:PerturbationsEx1}
\end{figure}

\begin{figure}[h!]
  \includegraphics[width=0.45\textwidth]{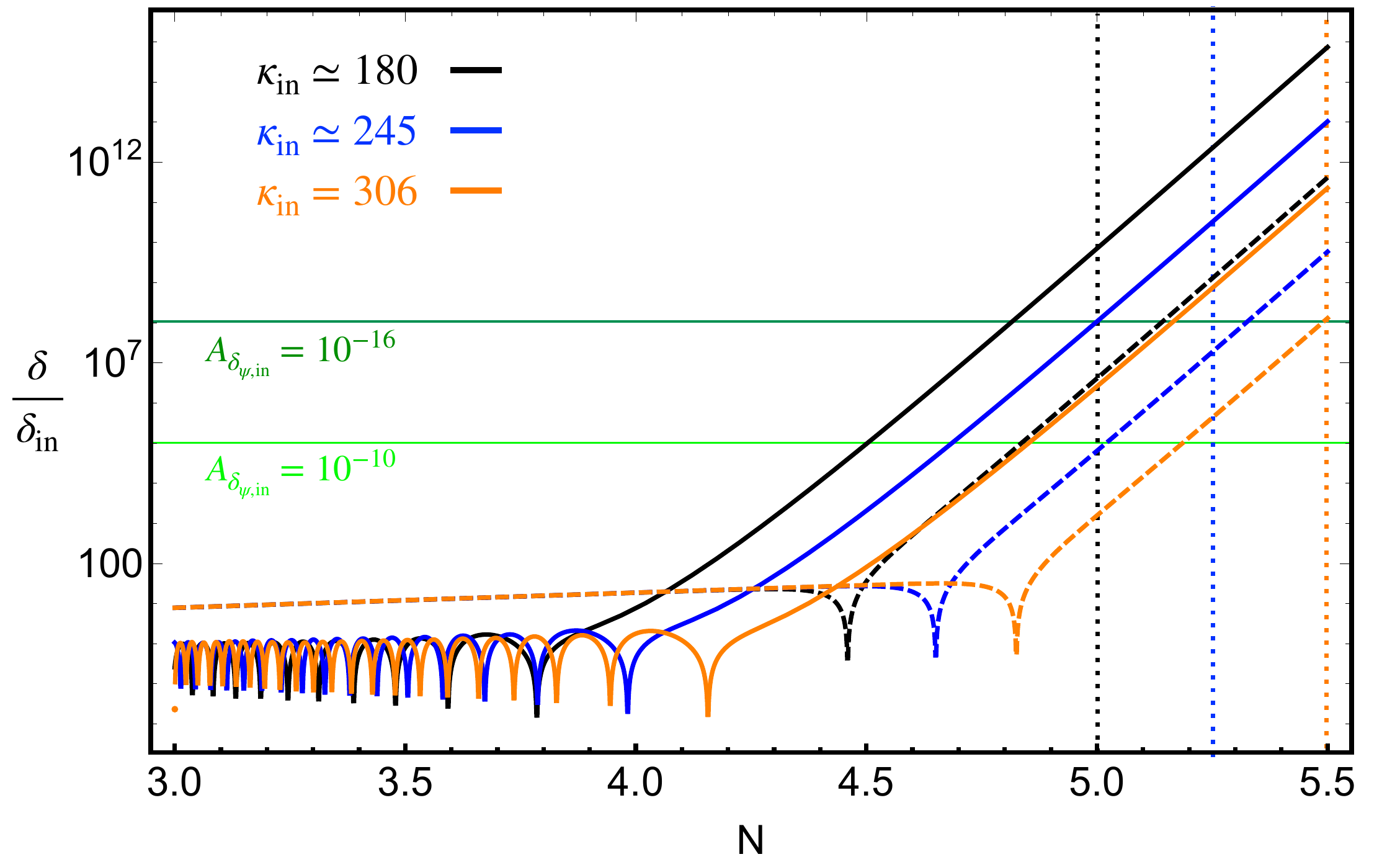}
\caption{Perturbations $\delta_\psi$ and $\delta_b$ for example 2. See the caption of Fig.~\ref{fig:PerturbationsEx1} for the legend.}
\label{fig:PerturbationsEx2}
\end{figure}

The equations governing the dynamics of the perturbations can be obtained in full generality by perturbing the Einstein equations as well as the conservation equations and the Klein-Gordon equation~\cite{Amendola:2003wa}. In our case, we will be interested in the perturbation of modes with wavelength $\lambda$ smaller than the Compton wavelength of the scalar field, namely $k/a \gg m_\phi$. In this limit, the equations of motion for the perturbations simplify. In Fourier space, the equations for the evolution of the fractional overdensities ($\delta_i$, $i=\psi, b$) and divergence of the dimensionless velocity perturbations ($\theta_i$, $i=\psi, b$) are\footnote{We follow the conventions of~\cite{Amendola:2003wa} for the definition of these.} 
\begin{align}
\label{eq:PerturbationDeltaPsi}
&\delta_\psi'' + \left[(w_\psi + 1) \theta_\psi\right]' + 3 \left[w_{\psi, \rho} (1-\beta \phi') \delta_\psi\right]' = 0 \,, \\
\label{eq:PerturbationDeltaB}
&\delta_b'' + \left(2 + \frac{H'}{H}\right) \delta_b' - \frac{3}{2} \Omega_\psi \delta_\psi - \frac{3}{2} \Omega_b \delta_b = 0 \,, 
\end{align}
where $w_{\psi, \rho} = \frac{d w}{d \log \rho} = \rho w_\psi'/\rho'$ and
\begin{align}
&\theta_\psi \simeq - \frac{1}{w_\psi+1} \left[\delta_\psi' + 3 w_{\psi, \rho} (1 - \beta \phi') \delta_\psi\right] \,, \nonumber\\
&\theta_\psi' \simeq - f \theta_\psi + \frac{w_\psi + w_{\psi, \rho}}{1 + w_\psi} \kappa^2 \delta_\psi - \omega_\psi \delta_\psi - \frac{3}{2} (1+w_\psi) \Omega_b \delta_b \,,\nonumber
\end{align}
where $\kappa = k/(a H)$ and
\begin{align}
& f = \left[(1-3 w_\psi)(1 - \beta \phi') - w_{\psi, \rho} A + 1 + \frac{H'}{H}\right] \theta_\psi \,, \nonumber\\
& A = 3 + \beta \phi' \frac{1-3w_\psi}{1+w_\psi} \,, \nonumber\\
&\text{\small $\omega_\psi = \frac{3}{2} (1+w_\psi) \Omega_\psi \left[1 + 2 \beta_\psi^2 \frac{(1-3w_\psi)(1-3w_\psi - 3 w_{\psi,\rho})}{(1+w_\psi)^2}\right] \,.$} \nonumber
\end{align}
We solve numerically this system of equation with ‘adiabiatic’ initial conditions $\delta_{\psi, \rm{in}} = \delta_{b, \rm{in}}$ and $(\delta'_{b, \rm{in}}, \delta'_{\psi, \rm{in}})=(0,0)$ for various\footnote{We do not consider smaller values of $\kappa$ than those reported in the legends of Fig.~\ref{fig:PerturbationsEx1} and Fig.~\ref{fig:PerturbationsEx2} because they do not go non-linear within the regime of validity of Eq.~\eqref{eq:PerturbationDeltaPsi} and Eq.~\eqref{eq:PerturbationDeltaB}.} values of $\kappa$ at $N = 0$, $\kappa_{\rm in}$. The solutions exhibit an exponentially fast growth as shown in Fig.~\ref{fig:PerturbationsEx1} and Fig.~\ref{fig:PerturbationsEx2} for example 1 and example 2 respectively. Since Eq.~\eqref{eq:PerturbationDeltaPsi} and Eq.~\eqref{eq:PerturbationDeltaB} are valid in the sub-Compton regime, the solution for the various modes is valid until $k/a \gtrsim m_\phi$. The maximum $N$ at which the solution is reliable is denoted by the dotted lines in Fig.~\ref{fig:PerturbationsEx1} and Fig.~\ref{fig:PerturbationsEx2}.

\begin{figure}[h!]
  \includegraphics[width=0.45\textwidth]{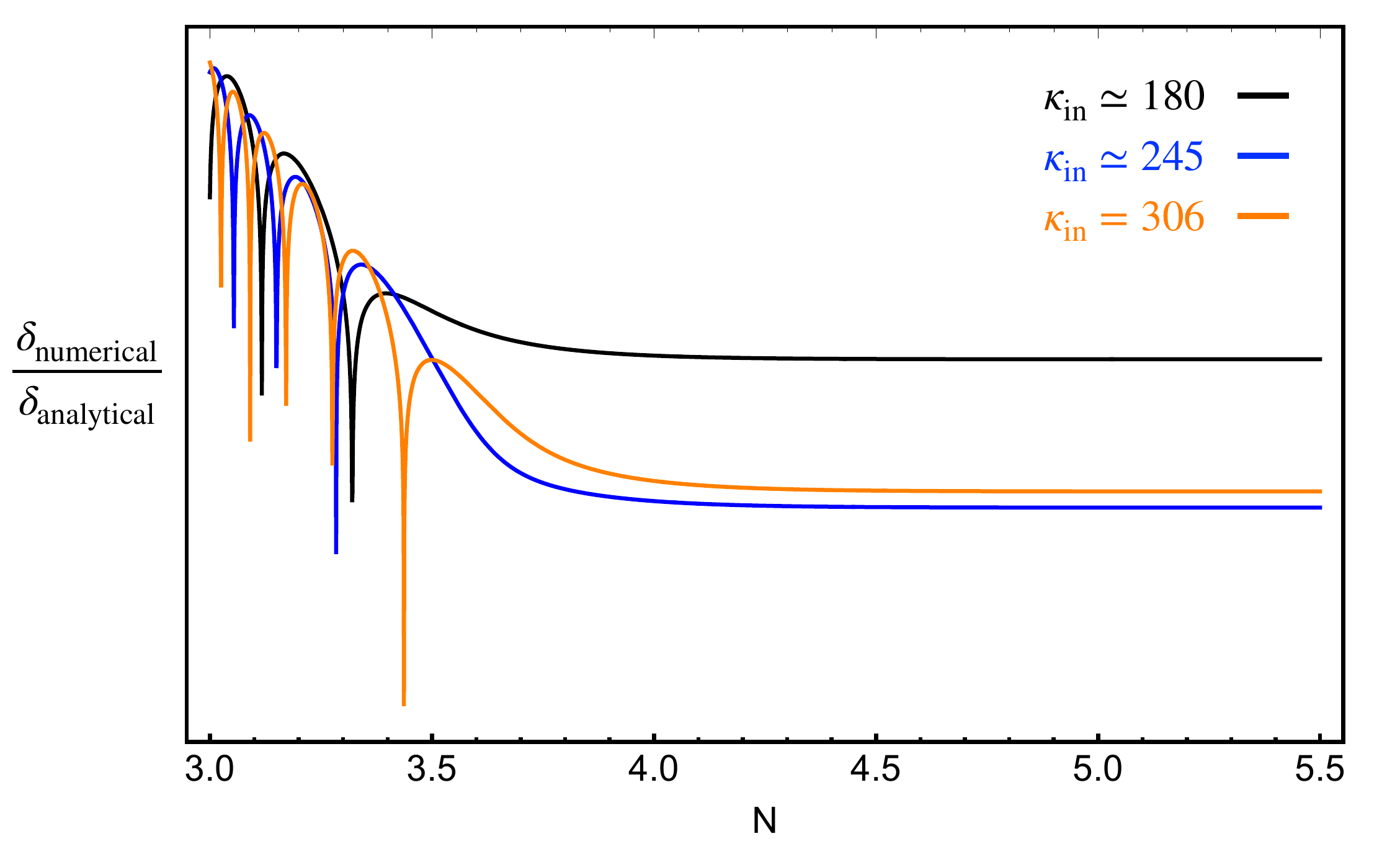}
\caption{Ratio between the numerical and analytical solutions for example 1. Asymptotically perturbations grow like $\delta_\psi \propto \exp(\gamma_\beta N)$, where $\gamma_\beta$ is given in Eq.~\eqref{eq:Exponent}.}
\label{fig:PerturbationsNorm}
\end{figure}

The system of equations has an exactly solvable regime. Once the $\psi$ component becomes non-relativistic, so that $w_\psi, w_{\rho, \psi} \simeq 0$, in the limit $\delta_{\psi} \gg \delta_b$ Eq.~\eqref{eq:PerturbationDeltaPsi} and Eq.~\eqref{eq:PerturbationDeltaB} simplify significantly:
\begin{align}
\label{eq:EqDeltaPsiSimplified}
&\delta_\psi'' + \frac{1}{2} \delta_\psi' - \tilde{\omega}_{\psi} \delta_\psi = 0 \,, \\
\label{eq:EqDeltaBSimplified}
&\delta_b'' + \frac{1}{2} \delta_b' - \frac{3}{2} \Omega_\psi \delta_\psi - \frac{3}{2} \Omega_b \delta_b = 0 \,,
\end{align}
where we have defined $\tilde{\omega}_\psi = \frac{3}{2} \Omega_\psi (1 + 2 \beta_\psi^2)$ and used the fact that $1 + \frac{H'}{H} \simeq \frac{1}{2}$ in a matter dominated universe.
Now, the equation for $\delta_\psi$ is decoupled and can be solved analytically. The growing mode is an exponential function: $\delta_\psi \propto \exp\left(\gamma_\beta N\right)$, with
\begin{equation}
\label{eq:Exponent}
\gamma_\beta = \frac{1}{4} \left(-1 + \sqrt{1 + 16 \tilde{\omega}_\psi}\right) \,.
\end{equation}
The exponents for the two examples under consideration are $\gamma_1 \simeq 13$ and $\gamma_2 \simeq 23$. Interestingly, in the regime $N \gtrsim N_{\rm NR}$, this is in agreement with the growth exponents obtained numerically (Fig.~\ref{fig:PerturbationsEx1} and Fig.~\ref{fig:PerturbationsEx2}) for adiabatic initial conditions. This indicates that the growth in $\delta_{\psi}$ essentially drives the growth in the system even for adiabatic initial conditions, see Fig.~\ref{fig:PerturbationsNorm}. We have checked this numerically. Note that in the scaling regime $w_\psi = w'_\psi = 0$, so the term proportional to $\kappa^2$ in Eq.~\eqref{eq:PerturbationDeltaPsi} and Eq.~\eqref{eq:PerturbationDeltaB} vanishes. However, when the change of the equation of state described by Eq.~\eqref{eq:EoS} is taken into account, it is important that this term goes to zero sufficiently fast, otherwise it would compete with the term proportional to $\tilde{\omega}_\psi$ in Eq.~\eqref{eq:EqDeltaPsiSimplified} that drives the exponential growth. We emphasise that it is in the scaling regime that Eq.~\eqref{eq:EqDeltaPsiSimplified} and Eq.~\eqref{eq:EqDeltaBSimplified} are valid.

The  solutions of Eq.~\eqref{eq:PerturbationDeltaPsi} and Eq.~\eqref{eq:PerturbationDeltaB} can be trusted until the linear approximation breaks down, namely when $\delta_{b} \big{/} \delta_{b, \rm in} \lesssim A^{-1/2}_{\delta_{b, \rm in}}$ and 
$\delta_\psi \big{/} \delta_{\psi, \rm in}  \lesssim  A^{-1/2}_{\delta_{\psi, \rm in}}$, where $A^{-1/2}_{\delta_{b, \rm in}}$ and  $A^{-1/2}_{\delta_{\psi, \rm in}}$  are the dimensionless strength of the perturbations as defined by the power spectra. We will use  $N_{\rm NL}$ to denote the e-foldings at which the validity of the linear theory breaks down and non-linearities become important. In general, $N_{\rm NL}$ will depend on the initial conditions $(\delta_{b, \rm{in}}, \delta_{\psi, \rm{in}})$. From Fig.~\ref{fig:PerturbationsEx1} and Fig.~\ref{fig:PerturbationsEx2} it is immediate to see the regions in which the linear approximation is valid: $N_{\rm NL}$ is fixed by the intersection of the perturbation mode curves with the horizontal light green (for $A_{\delta_{\psi, \rm in}} = 10^{-10}$) and dark green (for $A_{\delta_{\psi, \rm in}} = 10^{-16}$) lines. Note that for perturbations with both the reported initial amplitude values\footnote{Note that $A_{\delta_{\psi, \rm in}}=10^{-10}$ is the value expected from a scale invariant inflationary power spectrum.}, the modes go non-linear within very few e-foldings after the onset of the matter dominated epoch and well within the regime of validity of the equations for each mode. We note that this estimate is conservative as the perturbations can in principle undergo some growth before we begin to track them leading to a higher value
of $A_{\delta_{\psi, \rm in}}$. In both the examples and for the set of modes that we chose (which are among the first to go non-linear), we have $N_{\rm NL} \simeq 4.5$ for $A_{\delta_{\psi, \rm in}} = 10^{-10}$. We will use this value for the estimates presented in the following sections.
\subsection{Decays}
\label{sec:Decay}

To accommodate the successes of BBN, the matter dominated era has to end with a reheating temperature $T \gtrsim \text{a few MeV}$. This will happen if the widths of the three components $\Gamma_b$, $\Gamma_\psi$ and $\Gamma_\phi$ satisfy
\begin{equation}
\label{d1}
\Gamma_i \gtrsim H_{\rm BBN} \,, \quad i = b, \psi, \phi \,,
\end{equation}
where $H_{\rm BBN} \simeq 10^{-24} \, \text{GeV}$. We also require that the decays do not occur before the epoch of fast growth sets in and the perturbations grow rapidly. This implies:
\begin{equation}
\label{d2}
H_{\rm BBN} \lesssim \Gamma_{b, \phi, \psi} \lesssim H_{\rm NL} \,,
\end{equation}
where $H_{\rm NL}$ is the Hubble parameter at the time when the perturbations become non-linear.\\

If the background component is an oscillating modulus with mass $m_b$ and decay rate $\Gamma_b = m_b^3/\mpl^2$, then Eq.~\eqref{d2} would imply
\begin{equation}
m_{b, \rm min} \lesssim m_b \lesssim m_{b, \rm min} \exp\left(\frac{3}{4} (N_{\rm BBN} - N_{\rm NL})\right) \,,
\end{equation}
where $m_{b, \rm min} = \left(H_{\rm BBN} \mpl^2\right)^{1/3} \simeq 20 \, \text{TeV}$.

The conditions in Eq.~\eqref{d1} and Eq.~\eqref{d2} also constrain the strength of interaction of $\phi$ and $\psi$. For instance, let us consider the case in which the decays take place via Yukawa interactions
\begin{equation}
\mathcal{L}_{\rm int} \supset y \phi \bar{\chi} \chi + g \Psi \bar{\psi} \chi + \text{h.c.} \,,
\end{equation}
where $\chi$ is a visible sector fermion, while $\Psi$ is a visible sector scalar and $y$, $g$ are  Yukawa couplings. Then the decay rates are: 
\begin{equation}
\Gamma_\phi \simeq \frac{y^2 m_\phi}{8 \pi} \,, \quad \Gamma_{\psi} \simeq \frac{g^2 m_\psi}{16 \pi}.
\end{equation}
Concerning the $\psi$ component, the constraints in Eq.~\eqref{d2} translate into constraints for the product $g^2 m_\psi$. In the case of the scalar field $\phi$, since $m_\phi \simeq H_{\rm in}$ we can write the constraints in terms of $N_{\rm BBN}$ and $N_{\rm NL}$:
\begin{align}
y_{\rm min} \lesssim y \lesssim y_{\rm min} \, \exp\left(\frac{3}{4} (N_{\rm BBN} - N_{\rm NL})\right) \,,
\end{align}
where $N_{\rm BBN}$ is the value of $N$ at the time of BBN and
\begin{align}
&y_{\rm min} = \sqrt{8 \pi} \, \exp\left(-\frac{3}{4} N_{\rm BBN}\right) \,. \nonumber
\end{align}

The equations used in the evolution of the background and perturbations (in Sec.~\ref{sec:BackgroundDynamics} and the previous part of this section) do not incorporate the effects of the decays. They are in an instantaneous decay approximation and valid well before the decay processes play a significant role. Note that for a decay process with rate $\Gamma$, taking place in a matter dominated epoch, at times two e-foldings before $ t = \Gamma^{-1}$, the fraction of decayed particles is approximately five per cent. Thus, as a rule of thumb, we will require that the decay takes place at $N_{\rm dec}$, with $N_{\rm NL} \lesssim N_{\rm dec} \lesssim N_{\rm NL} + 2 \lesssim N_{\rm BBN}$.

For concreteness, taking $N_{\rm NR} = 2$ and $N_{\rm BBN} = N_{\rm NL} + 2$, we can give estimates for the Yukawa couplings. For instance, taking $N_{\rm NL} = 4.5$ from the numerical examples in Sec.~\ref{sec:Perturbations} yields $0.038 \lesssim y \lesssim 0.17$. 

\section{Phenomenological Implications}
\label{pheno}

  The fast growth of perturbations can have various interesting phenomenological implications. Here, we initiate their study. Understanding them in detail
so as to extract precise predictions requires detailed studies which we leave for future.

\subsection{Primordial Black Holes}
\label{sec:PBHs}

Once the perturbations become non-linear, it is reasonable to expect that the overdensities will collapse, forming either PBHs~\cite{1966AZh, 10.1093/mnras/152.1.75, Grindlay:1975eb, Chapline:1975ojl, Khlopov:1980mg, Polnarev:1985btg, Carr:2016drx} or other kinds of compact objects, such as oscillons (see e.g.~\cite{Antusch:2017flz} for a study of oscillon formation in the context of an EMD model), primordial halos~\cite{Savastano:2019zpr}, miniclusters~\cite{Hogan:1988mp, Fairbairn:2017sil} or star-like objects, see e.g.~\cite{Krippendorf:2018tei, Visinelli:2021uve} for two comprehensive reviews. The formation of PBHs in a matter dominated universe is on one hand facilitated by the fact that the background pressure vanishes~\cite{Harada:2016mhb}. On the other hand, though, any deviation from spherical symmetry will tend to virialise the collapsing system, avoiding the formation of a horizon. We defer a detailed numerical study of the formation of PBHs and microhalos, along the lines of~\cite{Helfer:2016ljl, Widdicombe:2018oeo, Muia:2019coe, Nazari:2020fmk, Eggemeier:2020zeg, Eggemeier:2021smj}, to a future work. In the present paper, we will provide simple estimates to exhibit the potentially rich phenomenology.\\

\begin{figure}[h!]
  \includegraphics[width=0.45\textwidth]{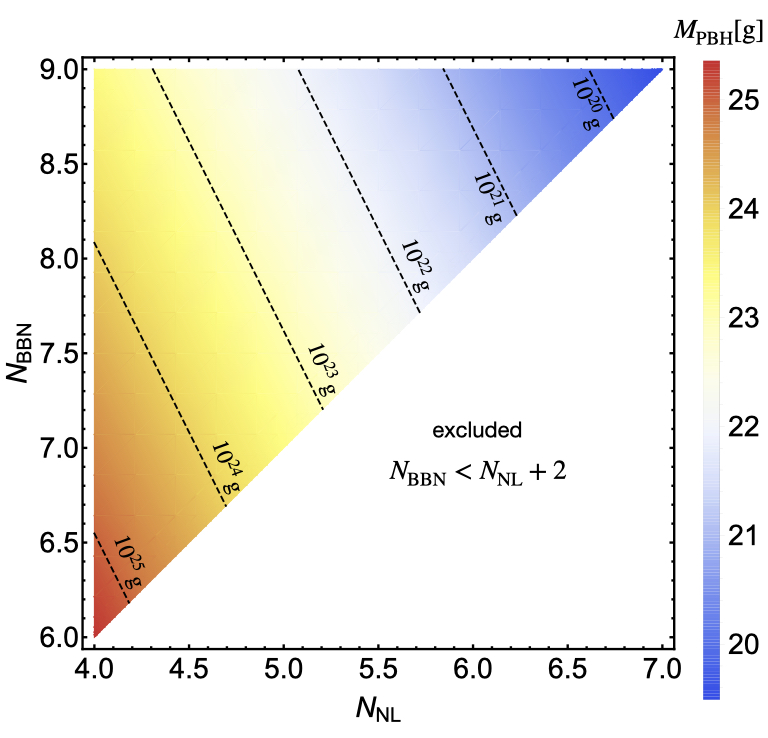}
\caption{Mass of the PBHs as a function of $N_{\rm NL}$ and $N_{\rm BBN}$.}
\label{fig:ParameterSpace}
\end{figure}

The easiest way to determine the mass scale of the PBHs that can be potentially formed is by isolating the scales for which the $\psi$ perturbations go non-linear~\cite{Amendola:2017xhl, Georg:2016yxa, Georg:2017mqk}. In our setup, the growth involves modes that are sub-Compton, i.e. $k/a > m_\phi$. For this reason, one can expect the maximum mass of the PBH formed to be
\begin{equation}
M_{\rm PBH} \simeq \rho(N_{\rm NL}) \times \left(\frac{\epsilon_m}{m_\phi}\right)^3 \,,
\label{eq:MPBH0}
\end{equation}
where $(\epsilon_m/m_\phi)$ with $\epsilon_m < 1$ parametrises the wavelength of the collapsing mode. We can estimate the various terms in this expression in terms of $N_{\rm NL}$ and $N_{\rm BBN}$. Assuming that the background is always matter dominated, we can write $H_{\rm NL}/H_{\rm in} \simeq \exp\left(- 3 N_{\rm NL}/2\right)$ and $H_{\rm BBN}/H_{\rm in} \simeq \exp\left(- 3 N_{\rm BBN}/2\right)$. Therefore, we find
\begin{equation}
\rho({N_{\rm NL}}) = 3 H_{\rm NL}^2 \mpl^2 = 3 \left(\frac{H_{\rm NL}}{H_{\rm in}}\right)^2 H_{\rm in}^2 \mpl^2 \,.
\end{equation}
Furthermore, we can approximate $H_{\rm in} \simeq m_\phi$ since the scalar field becomes dynamical at $N \simeq 0$. Hence, using for concreteness $\epsilon_m = 0.1$, we find from Eq.~\eqref{eq:MPBH0}
\begin{equation}
\label{eq:PBHMass}
M_{\rm PBH} \simeq 3 \times 10^{34} \, \text{g} \, \times  \exp\left(-\frac{3}{2} (2 N_{\rm NL} + N_{\rm BBN})\right) \,,
\end{equation}
where we have also used that $H_{\rm BBN} \simeq 10^{-24} \, \text{GeV}$. To make a concrete estimate, let us focus on the numbers that comes up from the numerical examples in Sec.~\ref{sec:Perturbations}. In those cases, perturbations with initial amplitude\footnote{We take the amplitude of the perturbations to be as given by the normalisation of scalar perturbations~\cite{Planck:2018vyg}, assuming a scale invariant power spectrum. $A_{\delta_{\psi, \rm{in}}} \simeq 10^{-10}$ enters the non-linear regime around $N_{\rm NL} = 4.5$ and $N_{\rm BBN} = N_{\rm NL} + 2 = 6.5$ (so that there is enough time for the various components to decay before the beginning of BBN, as explained in Sec.~\ref{sec:Decay}). In this case, Eq.~\eqref{eq:PBHMass} gives $M_{\rm PBH} \simeq 2.4 \times 10^{24} \, \text{g}$, which falls slightly above the sub-lunar mass range ($10^{17} \, {\rm g} \lesssim M_{\rm PBH} \lesssim 10^{23} \, {\rm g}$), in which PBHs can still compose $100 \%$ of dark matter~\cite{Carr:2021bzv}. Smaller values of $\beta$, smaller values for the initial amplitude of the perturbations or larger modes would give rise to larger values of $N_{\rm NL}$ and therefore slightly lighter PBHs, which would fall in the sub-lunar mass range. In Fig.~\ref{fig:ParameterSpace} we exhibit the PBH masses that can be obtained in our parameter space.}

Note that the expression in Eq.~\eqref{eq:MPBH0} gives a maximum value for the mass of the PBHs that can be formed. However, lighter PBHs can also be formed\footnote{The growth takes place for all modes for which  the term proportional to $\omega_{\psi}$ drives the dynamics in Eq.~\eqref{eq:PerturbationDeltaPsi} and Eq.~\eqref{eq:PerturbationDeltaB}.} PBHs with mass around $(10^{15}-10^{16}) \, \text{g}$ would be evaporating today and therefore are severely constrained from observations of the galactic and extra-galactic $\gamma$-rays background~\cite{Carr:2020gox}. Lighter PBHs, in the range $10^9 \, \rm{g} \lesssim M_{\rm PBH} \lesssim 10^{14} \, \rm{g}$ are subject to milder constraints due to BBN. Light PBHs, $M_{\rm PBH} \lesssim 10^{15} \, \rm{g}$ are quite interesting from the phenomenological point of view, as they might be a unique probe of the total number of light scalars in the fundamental theory~\cite{Calza:2021czr}, provide a baryogenesis mechanism~\cite{Hooper:2020otu}, reheat the universe~\cite{Lennon:2017tqq, Baldes:2020nuv} and produce GWs in the ultra-high-frequency band~\cite{Anantua:2008am, Dolgov:2011cq, Zagorac:2019ekv}.

Of course, to connect to phenomenology, one has to compute the fraction of PBH dark matter, i.e. $\beta = \rho_{\rm PBH}/\rho_{\rm DM}$, where $\rho_{\rm PBH}$ is the current dark matter energy density in PBHs, while $\rho_{\rm DM}$ is the current total dark matter energy density. Such a computation would require the knowledge of the threshold value $\delta_{\psi, \rm{c}}$ for a sub-Compton spherical overdensity $\delta_\psi$  for it to collapse to a PBH. In turn, computing $\delta_{\psi, \rm{c}}$ requires a careful numerical simulation that we plan to report in future work.

An early pre-BBN epoch of matter domination generically produces early micro- or mini-halos~\cite{ Blinov:2021axd, Barenboim:2021swl} due to the early growth of perturbations on scales below the horizon size. If these micro-halos are stable over cosmological time scale, their annihilation signature at present epoch from the dense galactic center has been studied extensively~\cite{Blanco:2019eij}. But in our case the situation is different: the halos made of $\psi$ particles give away scalar radiation. In fact, it has been shown that scalar radiation from early halos favours PBH formation~\cite{Flores:2020drq}. The remnant halos which do not form PBHs will be destroyed as the $\psi$ particles decay into the radiation bath of the Standard Model particles. So one naively expects that the number density of micro-halos will be very tiny at the present epoch unlike~\cite{Blinov:2021axd}.

\subsection{Gravitational Waves}
\label{sec:GWs}

Scalar perturbations generate GWs at second order in perturbation theory. This effect has been explored in several different contexts, see e.g.~\cite{Baumann:2007zm, Assadullahi:2009nf, Espinosa:2018eve, Kohri:2018awv, Inomata:2020yqv, Inomata:2020tkl, Domenech:2020ssp, Domenech:2021wkk} and~\cite{Domenech:2021ztg} for a recent comprehensive review on the subject. We can expect that this effect is also present in our scenario, as on sub-Compton scales the scalar field perturbations grow following the overdensity in the $\psi$ component~\cite{Amendola:2003wa}:
\begin{equation}
\label{eq:ScalarFieldMode}
\delta\phi_k \simeq \left(\frac{a H}{k}\right)^2 \beta_\psi \Omega_\psi \delta_\psi \,, \qquad \frac{k}{aH} \gg 1 \,.
\end{equation}
In this section we sketch some estimates of the amount of GWs and frequency that one can expect in the scenario described in the previous sections, due to second order scalar perturbations.\footnote{As we will only report order of magnitude estimates, we neglect factors containing the number of degrees of freedom in this section.} We leave a detailed computation of the GW spectrum to a future work and we follow~\cite{Giblin:2014gra, Chatrchyan:2020pzh} to do the estimates. Since we have a multi-component setup, second order scalar perturbations are not the only possible source of GWs. For instance, if the $\psi$ or the background components develop an anisotropic stress-energy tensor due to the rapid growth, they could also source GWs.\\

In order to do some estimates, let us do a few simplifying assumptions. We will assume that most of the GW production occurs at $N_{\rm NL}$, i.e. when the perturbations go non-linear and the scalar field fragments. We also assume that most of the GW energy is deposited in a mode $k_p/a \gtrsim m_\phi$, as it is reasonable to expect from Eq.~\eqref{eq:ScalarFieldMode}: at larger $k$ the scalar field perturbation is suppressed by the prefactor $(aH/k)$, while at lower $k$ the enhancement does not occur at all.\\

First, we would like to understand what is the typical frequency range that is involved. Given Eq.~\eqref{eq:ScalarFieldMode} we would expect that the signal is maximized when $\delta\phi_k$ is maximized. As modes are not amplified at $k/a \lesssim m_\phi$, the GW spectrum features a lower cutoff, given by the mass of the field, namely\footnote{A subscript ‘0’ denotes quantities evaluated at the present time.}
\begin{align}
\label{eq:F0}
f_0 \gtrsim \left(\frac{m_\phi}{10^{-17} \, \rm GeV}\right) \exp\left(N_{\rm NL} - N_{\rm BBN}\right) \times 10^{-3} \, \rm{Hz} \,,
\end{align}
where we have used that
\begin{align}
\label{eq:apa0}
\frac{a_{\rm NL}}{a_0} &= \frac{a_{\rm NL}}{a_{\rm BBN}} \frac{a_{\rm BBN}}{a_{\rm 0}} \simeq \\
&\simeq \exp\left[(N_{\rm NL} - N_{\rm BBN})\right] \times 10^{-10} \times \, \frac{\text{MeV}}{T_{\rm BBN}} \,. \nonumber
\end{align}
Eq.~\eqref{eq:F0} tells us that the signal could cover a large fraction of the GW spectrum depending on the mass $m_\phi \gtrsim 10^{-17} \, \rm{GeV}$ of the scalar field. Depending on the amplitude of the GW spectrum, the range of frequencies that is in principle involved could be probed by current and future GW experiments including LISA~\cite{amaroseoane2017laser, Barausse:2020rsu} ($f_0 \sim 10^{-2} \, \rm{Hz}$), DECIGO~\cite{Seto:2001qf} and BBO~\cite{Yagi:2011wg} ($f_0 \sim 1 \, \rm{Hz}$), LIGO/Virgo/KAGRA~\cite{KAGRA:2021kbb}, Einstein Telescope~\cite{Maggiore:2019uih} and Cosmic Explorer~\cite{Evans:2021gyd} ($f_0 \sim 100 \, \rm{Hz}$) and ultra-high-frequency band proposals~\cite{Aggarwal:2020olq} ($f_0 \gtrsim 10^3 \, \rm{Hz}$).\\

In order to estimate the amplitude of the GW spectrum let us parametrise\footnote{As in~\cite{Chatrchyan:2020pzh}, we neglect the tensor structure of the perturbation for the purposes of making a basic estimate.} $\Pi^{\rm TT}_{ij} \simeq \alpha \rho_\phi$, where $\Pi^{\rm TT}_{ij}$ is the transverse traceless component of the scalar field stress-energy tensor and $\alpha \lesssim 1$. Then one can write the peak fractional energy density in GWs at production ($N_{\rm NL}$) as~\cite{Chatrchyan:2020pzh}
\begin{align}
\label{eq:OmegaGWk}
\Omega_{{\rm GW}, p}(k_p) &\sim \frac{64 \pi^2}{3 \mpl^4 H_{\rm NL}^4} \frac{\rho_{\phi}^2(N_{\rm NL})}{(k_p/(a_{\rm NL} H_{\rm NL}))^2} \frac{\alpha^2}{\lambda} \nonumber \\
& \simeq \frac{192 \pi^2 \, \Omega_\phi^2(N_{\rm NL})}{(k_p/(a_{\rm NL} H_{\rm NL}))^2} \frac{\alpha^2}{\lambda} \,,
\end{align}
where $\lambda$ parametrizes the logritmic width of the signal $\lambda = \Delta \log k$. 
In order to compute $k_p/(a_{\rm NL} H_{\rm NL})$ we impose that the relevant scale is sub-Compton at $N_{\rm NL}$: $\frac{k_p}{a_{\rm NL}} \gtrsim m_\phi$, which implies
\begin{equation}
\frac{k_p}{a_{\rm NL} H_{\rm NL}} \gtrsim \exp\left(\frac{3 N_{\rm NL}}{2}\right) \,,
\end{equation}
where we have used that $m_\phi \sim H_{\rm in}$. Using $N_{\rm NL}= 4.5$ from the previous sections, one gets $k_p/(a_{\rm NL} H_{\rm NL}) \gtrsim 850$. Taking $\Omega_\phi(N_{\rm NL}) \simeq5 \times 10^{-3}$ from the numerics of the previous section and $\alpha \sim \lambda \sim 1$ one finds $\Omega_{{\rm GW}, p}(k_p) \simeq 6.5 \times 10^{-8}$.

The current fractional energy density in GWs can be computed by redshifting the result in Eq.~\eqref{eq:OmegaGWk} using Eq.~\eqref{eq:apa0}
\begin{align}
\Omega_{\rm GW, 0} (k_p) &= \left(\frac{a_{\rm NL}}{a_0}\right)^4 \frac{\rho_{\rm NL}}{\rho_0} \, \Omega_{{\rm GW}, p} (k_p) \simeq \\
& \simeq 10^{-4} e^{N_{\rm NL} - N_{\rm dec}} \, \Omega_{{\rm GW}, p}(k_p) \lesssim 2.4 \times 10^{-12} \,, \nonumber
\end{align}
where in the last step we have used the above estimate for $\Omega_{{\rm GW},p}(k_p)$ and $N_{\rm dec} - N_{\rm NL} = 1$.\\

Of course, in order to properly compute the the GW amplitude, we should track the behaviour of $h_k$ all the way through, starting from $N_{\rm NR}$ to the start of the standard radiation domination phase that begins when all the components $b$, $\phi$ and $\psi$ decay. This implies tracking perturbations also when they enter the non-linear regime, which would need a full numerical analysis. In this way one would be able to understand the effects of non-linearities~\cite{Delos:2020mtj, Delos:2019tsl,Delos:2018ueo} ad whether they would enhance the GW spectrum. We plan to further study GW production, including the numerical analysis, in a future work. Nevertheless, an amplitude $\Omega_{{\rm GW}, 0} \simeq 10^{-12}$ can be probed by most of the future GW detectors mentioned above (see~\cite{Thrane:2013oya, Mingarelli:2019mvk} for a detailed discussion).\\

Beyond the GW production mechanisms mentioned above, there are a couple of additional sources related to the formation of light PBHs:
\begin{itemize}[leftmargin=*]
\item Evaporation of light PBHs~\cite{Anantua:2008am, Dolgov:2011cq}, that produce a GW spectrum with peak at ultra-high-frequency, typically above  $10^{10} \, \rm{Hz}$. In~\cite{Dolgov:2011cq}, the maximum amplitude for the GW spectrum is computed to be $\Omega_{\rm GW} \simeq 10^{-7}$. In our scenario, it is likely that the maximum amplitude would be slightly smaller, due to an additional period of EMD before the $b$, $\phi$ and $\psi$ components decay.
\item Mergers of PBHs~\cite{Zagorac:2019ekv, Dolgov:2011cq}: in this case the frequency can be estimated as the \textit{Innermost Stable Circular Orbit} (ISCO) frequency, namely
\begin{equation}
f \simeq 10^{15} \, {\rm Hz} \left(\frac{10^{20} \, {\rm g}}{M_{\rm PBH}}\right) \,,
\end{equation}
where we have assumed an equal mass for the two merging PBHs. Note that for $M_{\rm PBH} \simeq 10^{20} \, \rm{g}$, which is the relevant mass range for this paper, the ISCO frequency roughly falls into the frequency range that will be accessible with axion experiments~\cite{Ejlli:2019bqj} like ALPS II~\cite{Bahre:2013ywa} and JURA~\cite{doi:10.1146/annurev-nucl-102014-022120}. In order to claim the detectability of such mergers an estimate of the number of expected events at a given distance is needed, that depends on the probability of forming a binary for such light PBHs. We will analyse these points carefully in a future work.
\end{itemize}

\section{Conclusions}
\label{conclude}

  The key result of this article is a scenario for fast growth of cosmological perturbations. At first, we present cosmological solutions which asymptote  to a matter dominated era in the early universe (prior to BBN). Density perturbations in this
matter dominated epoch grow very fast, the primary reason for this is a scalar mediated force between dark fermions. Examples with explicit (numerical) computations of the growth exponent have been presented in Sec.~\ref{sec:Perturbations}. The goal of this paper is to present the first
explicit examples, studies of the parameter space of the models and the exploration other related mechanisms for fast growth of perturbations will be carried out in a future work. We also took a phenomenological approach to the  change in the equation of state of the $\psi$ component.  An interesting future direction would be to present a microscopic description of such a transition of the $\psi$ equation of state and study the evolution of the perturbations in the setting. Fast growth of perturbations can potentially have a whole host of interesting phenomenological implications. We have outlined these in the context of primordial black holes and gravitational waves in Sec.~\ref{pheno}. Our estimates indicate that we can obtain PBHs in the sub-lunar window, that can in principle constitute $100\%$ of dark matter. Also, GWs can be expected in a wide range of frequencies, from $10^{-3}$ Hz to $10^{15}$ Hz, with amplitudes which are in the detectable range with future experiments. Extracting detailed predictions requires analysis which is beyond the scope of the present article. We plan to report on these in subsequent works.

\section*{Acknowledgments}
We thank Stefano Savastano and Luca Amendola for email exchange regarding the perturbation growth in coupled quintessence cosmology within scaling regime. AM is supported in part by the SERB, DST, Government of India by the grant MTR/2019/000267.  SD acknowledges SERB grant CRG/2019/006147.  FM is funded by a UKRI/EPSRC Stephen
Hawking fellowship, grant reference EP/T017279/1 and
partially supported by the STFC consolidated grant
ST/P000681/1.


\bibliography{mybib}

\begin{thebibliography}{92}%
\makeatletter
\providecommand \@ifxundefined [1]{%
 \@ifx{#1\undefined}
}%
\providecommand \@ifnum [1]{%
 \ifnum #1\expandafter \@firstoftwo
 \else \expandafter \@secondoftwo
 \fi
}%
\providecommand \@ifx [1]{%
 \ifx #1\expandafter \@firstoftwo
 \else \expandafter \@secondoftwo
 \fi
}%
\providecommand \natexlab [1]{#1}%
\providecommand \enquote  [1]{``#1''}%
\providecommand \bibnamefont  [1]{#1}%
\providecommand \bibfnamefont [1]{#1}%
\providecommand \citenamefont [1]{#1}%
\providecommand \href@noop [0]{\@secondoftwo}%
\providecommand \href [0]{\begingroup \@sanitize@url \@href}%
\providecommand \@href[1]{\@@startlink{#1}\@@href}%
\providecommand \@@href[1]{\endgroup#1\@@endlink}%
\providecommand \@sanitize@url [0]{\catcode `\\12\catcode `\$12\catcode
  `\&12\catcode `\#12\catcode `\^12\catcode `\_12\catcode `\%12\relax}%
\providecommand \@@startlink[1]{}%
\providecommand \@@endlink[0]{}%
\providecommand \url  [0]{\begingroup\@sanitize@url \@url }%
\providecommand \@url [1]{\endgroup\@href {#1}{\urlprefix }}%
\providecommand \urlprefix  [0]{URL }%
\providecommand \Eprint [0]{\href }%
\providecommand \doibase [0]{http://dx.doi.org/}%
\providecommand \selectlanguage [0]{\@gobble}%
\providecommand \bibinfo  [0]{\@secondoftwo}%
\providecommand \bibfield  [0]{\@secondoftwo}%
\providecommand \translation [1]{[#1]}%
\providecommand \BibitemOpen [0]{}%
\providecommand \bibitemStop [0]{}%
\providecommand \bibitemNoStop [0]{.\EOS\space}%
\providecommand \EOS [0]{\spacefactor3000\relax}%
\providecommand \BibitemShut  [1]{\csname bibitem#1\endcsname}%
\let\auto@bib@innerbib\@empty
\bibitem [{\citenamefont {Allahverdi}\ \emph {et~al.}(2020)\citenamefont
  {Allahverdi} \emph {et~al.}}]{Allahverdi:2020bys}%
  \BibitemOpen
  \bibfield  {author} {\bibinfo {author} {\bibfnamefont {R.}~\bibnamefont
  {Allahverdi}} \emph {et~al.},\ }\href {\doibase 10.21105/astro.2006.16182} {\
   (\bibinfo {year} {2020}),\ 10.21105/astro.2006.16182},\ \Eprint
  {http://arxiv.org/abs/2006.16182} {arXiv:2006.16182 [astro-ph.CO]}
  \BibitemShut {NoStop}%
\bibitem [{\citenamefont {Georg}\ \emph {et~al.}(2016)\citenamefont {Georg},
  \citenamefont {Sengor},\ and\ \citenamefont {Watson}}]{Georg:2016yxa}%
  \BibitemOpen
  \bibfield  {author} {\bibinfo {author} {\bibfnamefont {J.}~\bibnamefont
  {Georg}}, \bibinfo {author} {\bibfnamefont {G.}~\bibnamefont {Sengor}}, \
  and\ \bibinfo {author} {\bibfnamefont {S.}~\bibnamefont {Watson}},\ }\href
  {\doibase 10.1103/PhysRevD.93.123523} {\bibfield  {journal} {\bibinfo
  {journal} {Phys. Rev. D}\ }\textbf {\bibinfo {volume} {93}},\ \bibinfo
  {pages} {123523} (\bibinfo {year} {2016})},\ \Eprint
  {http://arxiv.org/abs/1603.00023} {arXiv:1603.00023 [hep-ph]} \BibitemShut
  {NoStop}%
\bibitem [{\citenamefont {Georg}\ and\ \citenamefont
  {Watson}(2017)}]{Georg:2017mqk}%
  \BibitemOpen
  \bibfield  {author} {\bibinfo {author} {\bibfnamefont {J.}~\bibnamefont
  {Georg}}\ and\ \bibinfo {author} {\bibfnamefont {S.}~\bibnamefont {Watson}},\
  }\href {\doibase 10.1007/JHEP09(2017)138} {\bibfield  {journal} {\bibinfo
  {journal} {JHEP}\ }\textbf {\bibinfo {volume} {09}},\ \bibinfo {pages} {138}
  (\bibinfo {year} {2017})},\ \Eprint {http://arxiv.org/abs/1703.04825}
  {arXiv:1703.04825 [astro-ph.CO]} \BibitemShut {NoStop}%
\bibitem [{\citenamefont {Erickcek}\ and\ \citenamefont
  {Sigurdson}(2011)}]{Erickcek:2011us}%
  \BibitemOpen
  \bibfield  {author} {\bibinfo {author} {\bibfnamefont {A.~L.}\ \bibnamefont
  {Erickcek}}\ and\ \bibinfo {author} {\bibfnamefont {K.}~\bibnamefont
  {Sigurdson}},\ }\href {\doibase 10.1103/PhysRevD.84.083503} {\bibfield
  {journal} {\bibinfo  {journal} {Phys. Rev. D}\ }\textbf {\bibinfo {volume}
  {84}},\ \bibinfo {pages} {083503} (\bibinfo {year} {2011})},\ \Eprint
  {http://arxiv.org/abs/1106.0536} {arXiv:1106.0536 [astro-ph.CO]} \BibitemShut
  {NoStop}%
\bibitem [{\citenamefont {Redmond}\ \emph {et~al.}(2018)\citenamefont
  {Redmond}, \citenamefont {Trezza},\ and\ \citenamefont
  {Erickcek}}]{Redmond:2018xty}%
  \BibitemOpen
  \bibfield  {author} {\bibinfo {author} {\bibfnamefont {K.}~\bibnamefont
  {Redmond}}, \bibinfo {author} {\bibfnamefont {A.}~\bibnamefont {Trezza}}, \
  and\ \bibinfo {author} {\bibfnamefont {A.~L.}\ \bibnamefont {Erickcek}},\
  }\href {\doibase 10.1103/PhysRevD.98.063504} {\bibfield  {journal} {\bibinfo
  {journal} {Phys. Rev. D}\ }\textbf {\bibinfo {volume} {98}},\ \bibinfo
  {pages} {063504} (\bibinfo {year} {2018})},\ \Eprint
  {http://arxiv.org/abs/1807.01327} {arXiv:1807.01327 [astro-ph.CO]}
  \BibitemShut {NoStop}%
\bibitem [{\citenamefont {Coughlan}\ \emph {et~al.}(1983)\citenamefont
  {Coughlan}, \citenamefont {Fischler}, \citenamefont {Kolb}, \citenamefont
  {Raby},\ and\ \citenamefont {Ross}}]{Coughlan:1983ci}%
  \BibitemOpen
  \bibfield  {author} {\bibinfo {author} {\bibfnamefont {G.~D.}\ \bibnamefont
  {Coughlan}}, \bibinfo {author} {\bibfnamefont {W.}~\bibnamefont {Fischler}},
  \bibinfo {author} {\bibfnamefont {E.~W.}\ \bibnamefont {Kolb}}, \bibinfo
  {author} {\bibfnamefont {S.}~\bibnamefont {Raby}}, \ and\ \bibinfo {author}
  {\bibfnamefont {G.~G.}\ \bibnamefont {Ross}},\ }\href {\doibase
  10.1016/0370-2693(83)91091-2} {\bibfield  {journal} {\bibinfo  {journal}
  {Phys. Lett. B}\ }\textbf {\bibinfo {volume} {131}},\ \bibinfo {pages} {59}
  (\bibinfo {year} {1983})}\BibitemShut {NoStop}%
\bibitem [{\citenamefont {Banks}\ \emph {et~al.}(1994)\citenamefont {Banks},
  \citenamefont {Kaplan},\ and\ \citenamefont {Nelson}}]{hep-ph/9308292}%
  \BibitemOpen
  \bibfield  {author} {\bibinfo {author} {\bibfnamefont {T.}~\bibnamefont
  {Banks}}, \bibinfo {author} {\bibfnamefont {D.~B.}\ \bibnamefont {Kaplan}}, \
  and\ \bibinfo {author} {\bibfnamefont {A.~E.}\ \bibnamefont {Nelson}},\
  }\href {\doibase 10.1103/PhysRevD.49.779} {\bibfield  {journal} {\bibinfo
  {journal} {Phys. Rev. D}\ }\textbf {\bibinfo {volume} {49}},\ \bibinfo
  {pages} {779} (\bibinfo {year} {1994})},\ \Eprint
  {http://arxiv.org/abs/hep-ph/9308292} {arXiv:hep-ph/9308292} \BibitemShut
  {NoStop}%
\bibitem [{\citenamefont {de~Carlos}\ \emph {et~al.}(1993)\citenamefont
  {de~Carlos}, \citenamefont {Casas}, \citenamefont {Quevedo},\ and\
  \citenamefont {Roulet}}]{hep-ph/9308325}%
  \BibitemOpen
  \bibfield  {author} {\bibinfo {author} {\bibfnamefont {B.}~\bibnamefont
  {de~Carlos}}, \bibinfo {author} {\bibfnamefont {J.~A.}\ \bibnamefont
  {Casas}}, \bibinfo {author} {\bibfnamefont {F.}~\bibnamefont {Quevedo}}, \
  and\ \bibinfo {author} {\bibfnamefont {E.}~\bibnamefont {Roulet}},\ }\href
  {\doibase 10.1016/0370-2693(93)91538-X} {\bibfield  {journal} {\bibinfo
  {journal} {Phys. Lett. B}\ }\textbf {\bibinfo {volume} {318}},\ \bibinfo
  {pages} {447} (\bibinfo {year} {1993})},\ \Eprint
  {http://arxiv.org/abs/hep-ph/9308325} {arXiv:hep-ph/9308325} \BibitemShut
  {NoStop}%
\bibitem [{\citenamefont {Dine}\ \emph {et~al.}(1996)\citenamefont {Dine},
  \citenamefont {Randall},\ and\ \citenamefont {Thomas}}]{hep-ph/9507453}%
  \BibitemOpen
  \bibfield  {author} {\bibinfo {author} {\bibfnamefont {M.}~\bibnamefont
  {Dine}}, \bibinfo {author} {\bibfnamefont {L.}~\bibnamefont {Randall}}, \
  and\ \bibinfo {author} {\bibfnamefont {S.~D.}\ \bibnamefont {Thomas}},\
  }\href {\doibase 10.1016/0550-3213(95)00538-2} {\bibfield  {journal}
  {\bibinfo  {journal} {Nucl. Phys. B}\ }\textbf {\bibinfo {volume} {458}},\
  \bibinfo {pages} {291} (\bibinfo {year} {1996})},\ \Eprint
  {http://arxiv.org/abs/hep-ph/9507453} {arXiv:hep-ph/9507453} \BibitemShut
  {NoStop}%
\bibitem [{\citenamefont {Cicoli}\ \emph {et~al.}(2016)\citenamefont {Cicoli},
  \citenamefont {Dutta}, \citenamefont {Maharana},\ and\ \citenamefont
  {Quevedo}}]{Cicoli:2016olq}%
  \BibitemOpen
  \bibfield  {author} {\bibinfo {author} {\bibfnamefont {M.}~\bibnamefont
  {Cicoli}}, \bibinfo {author} {\bibfnamefont {K.}~\bibnamefont {Dutta}},
  \bibinfo {author} {\bibfnamefont {A.}~\bibnamefont {Maharana}}, \ and\
  \bibinfo {author} {\bibfnamefont {F.}~\bibnamefont {Quevedo}},\ }\href
  {\doibase 10.1088/1475-7516/2016/08/006} {\bibfield  {journal} {\bibinfo
  {journal} {JCAP}\ }\textbf {\bibinfo {volume} {08}},\ \bibinfo {pages} {006}
  (\bibinfo {year} {2016})},\ \Eprint {http://arxiv.org/abs/1604.08512}
  {arXiv:1604.08512 [hep-th]} \BibitemShut {NoStop}%
\bibitem [{\citenamefont {Acharya}\ \emph
  {et~al.}(2019{\natexlab{a}})\citenamefont {Acharya}, \citenamefont {Dhuria},
  \citenamefont {Ghosh}, \citenamefont {Maharana},\ and\ \citenamefont
  {Muia}}]{1906.03025}%
  \BibitemOpen
  \bibfield  {author} {\bibinfo {author} {\bibfnamefont {B.~S.}\ \bibnamefont
  {Acharya}}, \bibinfo {author} {\bibfnamefont {M.}~\bibnamefont {Dhuria}},
  \bibinfo {author} {\bibfnamefont {D.}~\bibnamefont {Ghosh}}, \bibinfo
  {author} {\bibfnamefont {A.}~\bibnamefont {Maharana}}, \ and\ \bibinfo
  {author} {\bibfnamefont {F.}~\bibnamefont {Muia}},\ }\href {\doibase
  10.1088/1475-7516/2019/11/035} {\bibfield  {journal} {\bibinfo  {journal}
  {JCAP}\ }\textbf {\bibinfo {volume} {11}},\ \bibinfo {pages} {035} (\bibinfo
  {year} {2019}{\natexlab{a}})},\ \Eprint {http://arxiv.org/abs/1906.03025}
  {arXiv:1906.03025 [hep-th]} \BibitemShut {NoStop}%
\bibitem [{\citenamefont {Kane}\ \emph {et~al.}(2015)\citenamefont {Kane},
  \citenamefont {Sinha},\ and\ \citenamefont {Watson}}]{1502.07746}%
  \BibitemOpen
  \bibfield  {author} {\bibinfo {author} {\bibfnamefont {G.}~\bibnamefont
  {Kane}}, \bibinfo {author} {\bibfnamefont {K.}~\bibnamefont {Sinha}}, \ and\
  \bibinfo {author} {\bibfnamefont {S.}~\bibnamefont {Watson}},\ }\href
  {\doibase 10.1142/S0218271815300220} {\bibfield  {journal} {\bibinfo
  {journal} {Int. J. Mod. Phys. D}\ }\textbf {\bibinfo {volume} {24}},\
  \bibinfo {pages} {1530022} (\bibinfo {year} {2015})},\ \Eprint
  {http://arxiv.org/abs/1502.07746} {arXiv:1502.07746 [hep-th]} \BibitemShut
  {NoStop}%
\bibitem [{\citenamefont {Das}\ \emph {et~al.}(2006)\citenamefont {Das},
  \citenamefont {Corasaniti},\ and\ \citenamefont {Khoury}}]{Das:2005yj}%
  \BibitemOpen
  \bibfield  {author} {\bibinfo {author} {\bibfnamefont {S.}~\bibnamefont
  {Das}}, \bibinfo {author} {\bibfnamefont {P.~S.}\ \bibnamefont {Corasaniti}},
  \ and\ \bibinfo {author} {\bibfnamefont {J.}~\bibnamefont {Khoury}},\ }\href
  {\doibase 10.1103/PhysRevD.73.083509} {\bibfield  {journal} {\bibinfo
  {journal} {Phys. Rev. D}\ }\textbf {\bibinfo {volume} {73}},\ \bibinfo
  {pages} {083509} (\bibinfo {year} {2006})},\ \Eprint
  {http://arxiv.org/abs/astro-ph/0510628} {arXiv:astro-ph/0510628} \BibitemShut
  {NoStop}%
\bibitem [{\citenamefont {Amendola}(2000)}]{Amendola:1999er}%
  \BibitemOpen
  \bibfield  {author} {\bibinfo {author} {\bibfnamefont {L.}~\bibnamefont
  {Amendola}},\ }\href {\doibase 10.1103/PhysRevD.62.043511} {\bibfield
  {journal} {\bibinfo  {journal} {Phys. Rev. D}\ }\textbf {\bibinfo {volume}
  {62}},\ \bibinfo {pages} {043511} (\bibinfo {year} {2000})},\ \Eprint
  {http://arxiv.org/abs/astro-ph/9908023} {arXiv:astro-ph/9908023} \BibitemShut
  {NoStop}%
\bibitem [{\citenamefont {Vagnozzi}\ \emph {et~al.}(2021)\citenamefont
  {Vagnozzi}, \citenamefont {Visinelli}, \citenamefont {Brax}, \citenamefont
  {Davis},\ and\ \citenamefont {Sakstein}}]{Vagnozzi:2021quy}%
  \BibitemOpen
  \bibfield  {author} {\bibinfo {author} {\bibfnamefont {S.}~\bibnamefont
  {Vagnozzi}}, \bibinfo {author} {\bibfnamefont {L.}~\bibnamefont {Visinelli}},
  \bibinfo {author} {\bibfnamefont {P.}~\bibnamefont {Brax}}, \bibinfo {author}
  {\bibfnamefont {A.-C.}\ \bibnamefont {Davis}}, \ and\ \bibinfo {author}
  {\bibfnamefont {J.}~\bibnamefont {Sakstein}},\ }\href {\doibase
  10.1103/PhysRevD.104.063023} {\bibfield  {journal} {\bibinfo  {journal}
  {Phys. Rev. D}\ }\textbf {\bibinfo {volume} {104}},\ \bibinfo {pages}
  {063023} (\bibinfo {year} {2021})},\ \Eprint
  {http://arxiv.org/abs/2103.15834} {arXiv:2103.15834 [hep-ph]} \BibitemShut
  {NoStop}%
\bibitem [{\citenamefont {Tsai}\ \emph {et~al.}(2021)\citenamefont {Tsai},
  \citenamefont {Wu}, \citenamefont {Vagnozzi},\ and\ \citenamefont
  {Visinelli}}]{Tsai:2021irw}%
  \BibitemOpen
  \bibfield  {author} {\bibinfo {author} {\bibfnamefont {Y.-D.}\ \bibnamefont
  {Tsai}}, \bibinfo {author} {\bibfnamefont {Y.}~\bibnamefont {Wu}}, \bibinfo
  {author} {\bibfnamefont {S.}~\bibnamefont {Vagnozzi}}, \ and\ \bibinfo
  {author} {\bibfnamefont {L.}~\bibnamefont {Visinelli}},\ }\href@noop {} {\
  (\bibinfo {year} {2021})},\ \Eprint {http://arxiv.org/abs/2107.04038}
  {arXiv:2107.04038 [hep-ph]} \BibitemShut {NoStop}%
\bibitem [{\citenamefont {Savastano}\ \emph {et~al.}(2019)\citenamefont
  {Savastano}, \citenamefont {Amendola}, \citenamefont {Rubio},\ and\
  \citenamefont {Wetterich}}]{Savastano:2019zpr}%
  \BibitemOpen
  \bibfield  {author} {\bibinfo {author} {\bibfnamefont {S.}~\bibnamefont
  {Savastano}}, \bibinfo {author} {\bibfnamefont {L.}~\bibnamefont {Amendola}},
  \bibinfo {author} {\bibfnamefont {J.}~\bibnamefont {Rubio}}, \ and\ \bibinfo
  {author} {\bibfnamefont {C.}~\bibnamefont {Wetterich}},\ }\href {\doibase
  10.1103/PhysRevD.100.083518} {\bibfield  {journal} {\bibinfo  {journal}
  {Phys. Rev. D}\ }\textbf {\bibinfo {volume} {100}},\ \bibinfo {pages}
  {083518} (\bibinfo {year} {2019})},\ \Eprint
  {http://arxiv.org/abs/1906.05300} {arXiv:1906.05300 [astro-ph.CO]}
  \BibitemShut {NoStop}%
\bibitem [{\citenamefont {Amendola}\ \emph {et~al.}(2018)\citenamefont
  {Amendola}, \citenamefont {Rubio},\ and\ \citenamefont
  {Wetterich}}]{Amendola:2017xhl}%
  \BibitemOpen
  \bibfield  {author} {\bibinfo {author} {\bibfnamefont {L.}~\bibnamefont
  {Amendola}}, \bibinfo {author} {\bibfnamefont {J.}~\bibnamefont {Rubio}}, \
  and\ \bibinfo {author} {\bibfnamefont {C.}~\bibnamefont {Wetterich}},\ }\href
  {\doibase 10.1103/PhysRevD.97.081302} {\bibfield  {journal} {\bibinfo
  {journal} {Phys. Rev. D}\ }\textbf {\bibinfo {volume} {97}},\ \bibinfo
  {pages} {081302} (\bibinfo {year} {2018})},\ \Eprint
  {http://arxiv.org/abs/1711.09915} {arXiv:1711.09915 [astro-ph.CO]}
  \BibitemShut {NoStop}%
\bibitem [{\citenamefont {Damour}\ \emph {et~al.}(1990)\citenamefont {Damour},
  \citenamefont {Gibbons},\ and\ \citenamefont {Gundlach}}]{Damour:1990tw}%
  \BibitemOpen
  \bibfield  {author} {\bibinfo {author} {\bibfnamefont {T.}~\bibnamefont
  {Damour}}, \bibinfo {author} {\bibfnamefont {G.~W.}\ \bibnamefont {Gibbons}},
  \ and\ \bibinfo {author} {\bibfnamefont {C.}~\bibnamefont {Gundlach}},\
  }\href {\doibase 10.1103/PhysRevLett.64.123} {\bibfield  {journal} {\bibinfo
  {journal} {Phys. Rev. Lett.}\ }\textbf {\bibinfo {volume} {64}},\ \bibinfo
  {pages} {123} (\bibinfo {year} {1990})}\BibitemShut {NoStop}%
\bibitem [{\citenamefont {Wetterich}(1995)}]{hep-th/9408025}%
  \BibitemOpen
  \bibfield  {author} {\bibinfo {author} {\bibfnamefont {C.}~\bibnamefont
  {Wetterich}},\ }\href@noop {} {\bibfield  {journal} {\bibinfo  {journal}
  {Astron. Astrophys.}\ }\textbf {\bibinfo {volume} {301}},\ \bibinfo {pages}
  {321} (\bibinfo {year} {1995})},\ \Eprint
  {http://arxiv.org/abs/hep-th/9408025} {arXiv:hep-th/9408025} \BibitemShut
  {NoStop}%
\bibitem [{\citenamefont {Gasperini}\ \emph {et~al.}(2002)\citenamefont
  {Gasperini}, \citenamefont {Piazza},\ and\ \citenamefont
  {Veneziano}}]{gr-qc/0108016}%
  \BibitemOpen
  \bibfield  {author} {\bibinfo {author} {\bibfnamefont {M.}~\bibnamefont
  {Gasperini}}, \bibinfo {author} {\bibfnamefont {F.}~\bibnamefont {Piazza}}, \
  and\ \bibinfo {author} {\bibfnamefont {G.}~\bibnamefont {Veneziano}},\ }\href
  {\doibase 10.1103/PhysRevD.65.023508} {\bibfield  {journal} {\bibinfo
  {journal} {Phys. Rev. D}\ }\textbf {\bibinfo {volume} {65}},\ \bibinfo
  {pages} {023508} (\bibinfo {year} {2002})},\ \Eprint
  {http://arxiv.org/abs/gr-qc/0108016} {arXiv:gr-qc/0108016} \BibitemShut
  {NoStop}%
\bibitem [{\citenamefont {Chimento}\ \emph {et~al.}(2003)\citenamefont
  {Chimento}, \citenamefont {Jakubi}, \citenamefont {Pavon},\ and\
  \citenamefont {Zimdahl}}]{astro-ph/0303145}%
  \BibitemOpen
  \bibfield  {author} {\bibinfo {author} {\bibfnamefont {L.~P.}\ \bibnamefont
  {Chimento}}, \bibinfo {author} {\bibfnamefont {A.~S.}\ \bibnamefont
  {Jakubi}}, \bibinfo {author} {\bibfnamefont {D.}~\bibnamefont {Pavon}}, \
  and\ \bibinfo {author} {\bibfnamefont {W.}~\bibnamefont {Zimdahl}},\ }\href
  {\doibase 10.1103/PhysRevD.67.083513} {\bibfield  {journal} {\bibinfo
  {journal} {Phys. Rev. D}\ }\textbf {\bibinfo {volume} {67}},\ \bibinfo
  {pages} {083513} (\bibinfo {year} {2003})},\ \Eprint
  {http://arxiv.org/abs/astro-ph/0303145} {arXiv:astro-ph/0303145} \BibitemShut
  {NoStop}%
\bibitem [{\citenamefont {Amendola}\ \emph {et~al.}(2003)\citenamefont
  {Amendola}, \citenamefont {Gasperini}, \citenamefont {Tocchini-Valentini},\
  and\ \citenamefont {Ungarelli}}]{astro-ph/0208032}%
  \BibitemOpen
  \bibfield  {author} {\bibinfo {author} {\bibfnamefont {L.}~\bibnamefont
  {Amendola}}, \bibinfo {author} {\bibfnamefont {M.}~\bibnamefont {Gasperini}},
  \bibinfo {author} {\bibfnamefont {D.}~\bibnamefont {Tocchini-Valentini}}, \
  and\ \bibinfo {author} {\bibfnamefont {C.}~\bibnamefont {Ungarelli}},\ }\href
  {\doibase 10.1103/PhysRevD.67.043512} {\bibfield  {journal} {\bibinfo
  {journal} {Phys. Rev. D}\ }\textbf {\bibinfo {volume} {67}},\ \bibinfo
  {pages} {043512} (\bibinfo {year} {2003})},\ \Eprint
  {http://arxiv.org/abs/astro-ph/0208032} {arXiv:astro-ph/0208032} \BibitemShut
  {NoStop}%
\bibitem [{\citenamefont {Rhodes}\ \emph {et~al.}(2003)\citenamefont {Rhodes},
  \citenamefont {van~de Bruck}, \citenamefont {Brax},\ and\ \citenamefont
  {Davis}}]{astro-ph/0306343}%
  \BibitemOpen
  \bibfield  {author} {\bibinfo {author} {\bibfnamefont {C.~S.}\ \bibnamefont
  {Rhodes}}, \bibinfo {author} {\bibfnamefont {C.}~\bibnamefont {van~de
  Bruck}}, \bibinfo {author} {\bibfnamefont {P.}~\bibnamefont {Brax}}, \ and\
  \bibinfo {author} {\bibfnamefont {A.~C.}\ \bibnamefont {Davis}},\ }\href
  {\doibase 10.1103/PhysRevD.68.083511} {\bibfield  {journal} {\bibinfo
  {journal} {Phys. Rev. D}\ }\textbf {\bibinfo {volume} {68}},\ \bibinfo
  {pages} {083511} (\bibinfo {year} {2003})},\ \Eprint
  {http://arxiv.org/abs/astro-ph/0306343} {arXiv:astro-ph/0306343} \BibitemShut
  {NoStop}%
\bibitem [{\citenamefont {Mangano}\ \emph {et~al.}(2003)\citenamefont
  {Mangano}, \citenamefont {Miele},\ and\ \citenamefont
  {Pettorino}}]{astro-ph/0212518}%
  \BibitemOpen
  \bibfield  {author} {\bibinfo {author} {\bibfnamefont {G.}~\bibnamefont
  {Mangano}}, \bibinfo {author} {\bibfnamefont {G.}~\bibnamefont {Miele}}, \
  and\ \bibinfo {author} {\bibfnamefont {V.}~\bibnamefont {Pettorino}},\ }\href
  {\doibase 10.1142/S0217732303009940} {\bibfield  {journal} {\bibinfo
  {journal} {Mod. Phys. Lett. A}\ }\textbf {\bibinfo {volume} {18}},\ \bibinfo
  {pages} {831} (\bibinfo {year} {2003})},\ \Eprint
  {http://arxiv.org/abs/astro-ph/0212518} {arXiv:astro-ph/0212518} \BibitemShut
  {NoStop}%
\bibitem [{\citenamefont {Hoffman}(2003)}]{astro-ph/0307350}%
  \BibitemOpen
  \bibfield  {author} {\bibinfo {author} {\bibfnamefont {M.~B.}\ \bibnamefont
  {Hoffman}},\ }\href@noop {} {\  (\bibinfo {year} {2003})},\ \Eprint
  {http://arxiv.org/abs/astro-ph/0307350} {arXiv:astro-ph/0307350} \BibitemShut
  {NoStop}%
\bibitem [{\citenamefont {Amendola}(2004)}]{Amendola:2003wa}%
  \BibitemOpen
  \bibfield  {author} {\bibinfo {author} {\bibfnamefont {L.}~\bibnamefont
  {Amendola}},\ }\href {\doibase 10.1103/PhysRevD.69.103524} {\bibfield
  {journal} {\bibinfo  {journal} {Phys. Rev. D}\ }\textbf {\bibinfo {volume}
  {69}},\ \bibinfo {pages} {103524} (\bibinfo {year} {2004})},\ \Eprint
  {http://arxiv.org/abs/astro-ph/0311175} {arXiv:astro-ph/0311175} \BibitemShut
  {NoStop}%
\bibitem [{\citenamefont {Dom\`enech}\ and\ \citenamefont
  {Sasaki}(2021)}]{Domenech:2021uyx}%
  \BibitemOpen
  \bibfield  {author} {\bibinfo {author} {\bibfnamefont {G.}~\bibnamefont
  {Dom\`enech}}\ and\ \bibinfo {author} {\bibfnamefont {M.}~\bibnamefont
  {Sasaki}},\ }\href {\doibase 10.1088/1475-7516/2021/06/030} {\bibfield
  {journal} {\bibinfo  {journal} {JCAP}\ }\textbf {\bibinfo {volume} {06}},\
  \bibinfo {pages} {030} (\bibinfo {year} {2021})},\ \Eprint
  {http://arxiv.org/abs/2104.05271} {arXiv:2104.05271 [hep-th]} \BibitemShut
  {NoStop}%
\bibitem [{\citenamefont {Gehrlein}\ and\ \citenamefont
  {Pierre}(2020)}]{Gehrlein:2019iwl}%
  \BibitemOpen
  \bibfield  {author} {\bibinfo {author} {\bibfnamefont {J.}~\bibnamefont
  {Gehrlein}}\ and\ \bibinfo {author} {\bibfnamefont {M.}~\bibnamefont
  {Pierre}},\ }\href {\doibase 10.1007/JHEP02(2020)068} {\bibfield  {journal}
  {\bibinfo  {journal} {JHEP}\ }\textbf {\bibinfo {volume} {02}},\ \bibinfo
  {pages} {068} (\bibinfo {year} {2020})},\ \Eprint
  {http://arxiv.org/abs/1912.06661} {arXiv:1912.06661 [hep-ph]} \BibitemShut
  {NoStop}%
\bibitem [{\citenamefont {Shelton}\ and\ \citenamefont
  {Zurek}(2010)}]{Shelton:2010ta}%
  \BibitemOpen
  \bibfield  {author} {\bibinfo {author} {\bibfnamefont {J.}~\bibnamefont
  {Shelton}}\ and\ \bibinfo {author} {\bibfnamefont {K.~M.}\ \bibnamefont
  {Zurek}},\ }\href {\doibase 10.1103/PhysRevD.82.123512} {\bibfield  {journal}
  {\bibinfo  {journal} {Phys. Rev. D}\ }\textbf {\bibinfo {volume} {82}},\
  \bibinfo {pages} {123512} (\bibinfo {year} {2010})},\ \Eprint
  {http://arxiv.org/abs/1008.1997} {arXiv:1008.1997 [hep-ph]} \BibitemShut
  {NoStop}%
\bibitem [{\citenamefont {Hinterbichler}\ and\ \citenamefont
  {Khoury}(2010)}]{Hinterbichler:2010es}%
  \BibitemOpen
  \bibfield  {author} {\bibinfo {author} {\bibfnamefont {K.}~\bibnamefont
  {Hinterbichler}}\ and\ \bibinfo {author} {\bibfnamefont {J.}~\bibnamefont
  {Khoury}},\ }\href {\doibase 10.1103/PhysRevLett.104.231301} {\bibfield
  {journal} {\bibinfo  {journal} {Phys. Rev. Lett.}\ }\textbf {\bibinfo
  {volume} {104}},\ \bibinfo {pages} {231301} (\bibinfo {year} {2010})},\
  \Eprint {http://arxiv.org/abs/1001.4525} {arXiv:1001.4525 [hep-th]}
  \BibitemShut {NoStop}%
\bibitem [{\citenamefont {Acharya}\ \emph
  {et~al.}(2019{\natexlab{b}})\citenamefont {Acharya}, \citenamefont
  {Maharana},\ and\ \citenamefont {Muia}}]{Acharya:2018deu}%
  \BibitemOpen
  \bibfield  {author} {\bibinfo {author} {\bibfnamefont {B.~S.}\ \bibnamefont
  {Acharya}}, \bibinfo {author} {\bibfnamefont {A.}~\bibnamefont {Maharana}}, \
  and\ \bibinfo {author} {\bibfnamefont {F.}~\bibnamefont {Muia}},\ }\href
  {\doibase 10.1007/JHEP03(2019)048} {\bibfield  {journal} {\bibinfo  {journal}
  {JHEP}\ }\textbf {\bibinfo {volume} {03}},\ \bibinfo {pages} {048} (\bibinfo
  {year} {2019}{\natexlab{b}})},\ \Eprint {http://arxiv.org/abs/1811.10633}
  {arXiv:1811.10633 [hep-th]} \BibitemShut {NoStop}%
\bibitem [{\citenamefont {{Zel'dovich}}\ and\ \citenamefont
  {{Novikov}}(1966)}]{1966AZh}%
  \BibitemOpen
  \bibfield  {author} {\bibinfo {author} {\bibfnamefont {Y.~B.}\ \bibnamefont
  {{Zel'dovich}}}\ and\ \bibinfo {author} {\bibfnamefont {I.~D.}\ \bibnamefont
  {{Novikov}}},\ }\href@noop {} {\bibfield  {journal} {\bibinfo  {journal}
  {Soviet Astronomy, Vol. 10, p.602}\ }\textbf {\bibinfo {volume} {43}},\
  \bibinfo {pages} {758} (\bibinfo {year} {1966})}\BibitemShut {NoStop}%
\bibitem [{\citenamefont {Hawking}(1971)}]{10.1093/mnras/152.1.75}%
  \BibitemOpen
  \bibfield  {author} {\bibinfo {author} {\bibfnamefont {S.}~\bibnamefont
  {Hawking}},\ }\href {\doibase 10.1093/mnras/152.1.75} {\bibfield  {journal}
  {\bibinfo  {journal} {Monthly Notices of the Royal Astronomical Society}\
  }\textbf {\bibinfo {volume} {152}},\ \bibinfo {pages} {75} (\bibinfo {year}
  {1971})},\ \Eprint
  {http://arxiv.org/abs/https://academic.oup.com/mnras/article-pdf/152/1/75/9360899/mnras152-0075.pdf}
  {https://academic.oup.com/mnras/article-pdf/152/1/75/9360899/mnras152-0075.pdf}
  \BibitemShut {NoStop}%
\bibitem [{\citenamefont {Grindlay}\ \emph {et~al.}(1975)\citenamefont
  {Grindlay}, \citenamefont {Helmken}, \citenamefont {Brown}, \citenamefont
  {Davis},\ and\ \citenamefont {Allen}}]{Grindlay:1975eb}%
  \BibitemOpen
  \bibfield  {author} {\bibinfo {author} {\bibfnamefont {J.~E.}\ \bibnamefont
  {Grindlay}}, \bibinfo {author} {\bibfnamefont {H.~F.}\ \bibnamefont
  {Helmken}}, \bibinfo {author} {\bibfnamefont {R.~H.}\ \bibnamefont {Brown}},
  \bibinfo {author} {\bibfnamefont {J.}~\bibnamefont {Davis}}, \ and\ \bibinfo
  {author} {\bibfnamefont {L.~R.}\ \bibnamefont {Allen}},\ }\href {\doibase
  10.1086/153861} {\bibfield  {journal} {\bibinfo  {journal} {Astrophys. J.}\
  }\textbf {\bibinfo {volume} {201}},\ \bibinfo {pages} {82} (\bibinfo {year}
  {1975})}\BibitemShut {NoStop}%
\bibitem [{\citenamefont {Chapline}(1975)}]{Chapline:1975ojl}%
  \BibitemOpen
  \bibfield  {author} {\bibinfo {author} {\bibfnamefont {G.~F.}\ \bibnamefont
  {Chapline}},\ }\href {\doibase 10.1038/253251a0} {\bibfield  {journal}
  {\bibinfo  {journal} {Nature}\ }\textbf {\bibinfo {volume} {253}},\ \bibinfo
  {pages} {251} (\bibinfo {year} {1975})}\BibitemShut {NoStop}%
\bibitem [{\citenamefont {Khlopov}\ and\ \citenamefont
  {Polnarev}(1980)}]{Khlopov:1980mg}%
  \BibitemOpen
  \bibfield  {author} {\bibinfo {author} {\bibfnamefont {M.~Y.}\ \bibnamefont
  {Khlopov}}\ and\ \bibinfo {author} {\bibfnamefont {A.~G.}\ \bibnamefont
  {Polnarev}},\ }\href {\doibase 10.1016/0370-2693(80)90624-3} {\bibfield
  {journal} {\bibinfo  {journal} {Phys. Lett. B}\ }\textbf {\bibinfo {volume}
  {97}},\ \bibinfo {pages} {383} (\bibinfo {year} {1980})}\BibitemShut
  {NoStop}%
\bibitem [{\citenamefont {Polnarev}\ and\ \citenamefont
  {Khlopov}(1985)}]{Polnarev:1985btg}%
  \BibitemOpen
  \bibfield  {author} {\bibinfo {author} {\bibfnamefont {A.~G.}\ \bibnamefont
  {Polnarev}}\ and\ \bibinfo {author} {\bibfnamefont {M.~Y.}\ \bibnamefont
  {Khlopov}},\ }\href {\doibase 10.1070/PU1985v028n03ABEH003858} {\bibfield
  {journal} {\bibinfo  {journal} {Sov. Phys. Usp.}\ }\textbf {\bibinfo {volume}
  {28}},\ \bibinfo {pages} {213} (\bibinfo {year} {1985})}\BibitemShut
  {NoStop}%
\bibitem [{\citenamefont {Carr}\ \emph {et~al.}(2016)\citenamefont {Carr},
  \citenamefont {Kuhnel},\ and\ \citenamefont {Sandstad}}]{Carr:2016drx}%
  \BibitemOpen
  \bibfield  {author} {\bibinfo {author} {\bibfnamefont {B.}~\bibnamefont
  {Carr}}, \bibinfo {author} {\bibfnamefont {F.}~\bibnamefont {Kuhnel}}, \ and\
  \bibinfo {author} {\bibfnamefont {M.}~\bibnamefont {Sandstad}},\ }\href
  {\doibase 10.1103/PhysRevD.94.083504} {\bibfield  {journal} {\bibinfo
  {journal} {Phys. Rev. D}\ }\textbf {\bibinfo {volume} {94}},\ \bibinfo
  {pages} {083504} (\bibinfo {year} {2016})},\ \Eprint
  {http://arxiv.org/abs/1607.06077} {arXiv:1607.06077 [astro-ph.CO]}
  \BibitemShut {NoStop}%
\bibitem [{\citenamefont {Antusch}\ \emph {et~al.}(2018)\citenamefont
  {Antusch}, \citenamefont {Cefala}, \citenamefont {Krippendorf}, \citenamefont
  {Muia}, \citenamefont {Orani},\ and\ \citenamefont
  {Quevedo}}]{Antusch:2017flz}%
  \BibitemOpen
  \bibfield  {author} {\bibinfo {author} {\bibfnamefont {S.}~\bibnamefont
  {Antusch}}, \bibinfo {author} {\bibfnamefont {F.}~\bibnamefont {Cefala}},
  \bibinfo {author} {\bibfnamefont {S.}~\bibnamefont {Krippendorf}}, \bibinfo
  {author} {\bibfnamefont {F.}~\bibnamefont {Muia}}, \bibinfo {author}
  {\bibfnamefont {S.}~\bibnamefont {Orani}}, \ and\ \bibinfo {author}
  {\bibfnamefont {F.}~\bibnamefont {Quevedo}},\ }\href {\doibase
  10.1007/JHEP01(2018)083} {\bibfield  {journal} {\bibinfo  {journal} {JHEP}\
  }\textbf {\bibinfo {volume} {01}},\ \bibinfo {pages} {083} (\bibinfo {year}
  {2018})},\ \Eprint {http://arxiv.org/abs/1708.08922} {arXiv:1708.08922
  [hep-th]} \BibitemShut {NoStop}%
\bibitem [{\citenamefont {Hogan}\ and\ \citenamefont
  {Rees}(1988)}]{Hogan:1988mp}%
  \BibitemOpen
  \bibfield  {author} {\bibinfo {author} {\bibfnamefont {C.~J.}\ \bibnamefont
  {Hogan}}\ and\ \bibinfo {author} {\bibfnamefont {M.~J.}\ \bibnamefont
  {Rees}},\ }\href {\doibase 10.1016/0370-2693(88)91655-3} {\bibfield
  {journal} {\bibinfo  {journal} {Phys. Lett. B}\ }\textbf {\bibinfo {volume}
  {205}},\ \bibinfo {pages} {228} (\bibinfo {year} {1988})}\BibitemShut
  {NoStop}%
\bibitem [{\citenamefont {Fairbairn}\ \emph {et~al.}(2018)\citenamefont
  {Fairbairn}, \citenamefont {Marsh}, \citenamefont {Quevillon},\ and\
  \citenamefont {Rozier}}]{Fairbairn:2017sil}%
  \BibitemOpen
  \bibfield  {author} {\bibinfo {author} {\bibfnamefont {M.}~\bibnamefont
  {Fairbairn}}, \bibinfo {author} {\bibfnamefont {D.~J.~E.}\ \bibnamefont
  {Marsh}}, \bibinfo {author} {\bibfnamefont {J.}~\bibnamefont {Quevillon}}, \
  and\ \bibinfo {author} {\bibfnamefont {S.}~\bibnamefont {Rozier}},\ }\href
  {\doibase 10.1103/PhysRevD.97.083502} {\bibfield  {journal} {\bibinfo
  {journal} {Phys. Rev. D}\ }\textbf {\bibinfo {volume} {97}},\ \bibinfo
  {pages} {083502} (\bibinfo {year} {2018})},\ \Eprint
  {http://arxiv.org/abs/1707.03310} {arXiv:1707.03310 [astro-ph.CO]}
  \BibitemShut {NoStop}%
\bibitem [{\citenamefont {Krippendorf}\ \emph {et~al.}(2018)\citenamefont
  {Krippendorf}, \citenamefont {Muia},\ and\ \citenamefont
  {Quevedo}}]{Krippendorf:2018tei}%
  \BibitemOpen
  \bibfield  {author} {\bibinfo {author} {\bibfnamefont {S.}~\bibnamefont
  {Krippendorf}}, \bibinfo {author} {\bibfnamefont {F.}~\bibnamefont {Muia}}, \
  and\ \bibinfo {author} {\bibfnamefont {F.}~\bibnamefont {Quevedo}},\ }\href
  {\doibase 10.1007/JHEP08(2018)070} {\bibfield  {journal} {\bibinfo  {journal}
  {JHEP}\ }\textbf {\bibinfo {volume} {08}},\ \bibinfo {pages} {070} (\bibinfo
  {year} {2018})},\ \Eprint {http://arxiv.org/abs/1806.04690} {arXiv:1806.04690
  [hep-th]} \BibitemShut {NoStop}%
\bibitem [{\citenamefont {Visinelli}(2021)}]{Visinelli:2021uve}%
  \BibitemOpen
  \bibfield  {author} {\bibinfo {author} {\bibfnamefont {L.}~\bibnamefont
  {Visinelli}},\ }\href {\doibase 10.1142/S0218271821300068} {\  (\bibinfo
  {year} {2021}),\ 10.1142/S0218271821300068},\ \Eprint
  {http://arxiv.org/abs/2109.05481} {arXiv:2109.05481 [gr-qc]} \BibitemShut
  {NoStop}%
\bibitem [{\citenamefont {Harada}\ \emph {et~al.}(2016)\citenamefont {Harada},
  \citenamefont {Yoo}, \citenamefont {Kohri}, \citenamefont {Nakao},\ and\
  \citenamefont {Jhingan}}]{Harada:2016mhb}%
  \BibitemOpen
  \bibfield  {author} {\bibinfo {author} {\bibfnamefont {T.}~\bibnamefont
  {Harada}}, \bibinfo {author} {\bibfnamefont {C.-M.}\ \bibnamefont {Yoo}},
  \bibinfo {author} {\bibfnamefont {K.}~\bibnamefont {Kohri}}, \bibinfo
  {author} {\bibfnamefont {K.-i.}\ \bibnamefont {Nakao}}, \ and\ \bibinfo
  {author} {\bibfnamefont {S.}~\bibnamefont {Jhingan}},\ }\href {\doibase
  10.3847/1538-4357/833/1/61} {\bibfield  {journal} {\bibinfo  {journal}
  {Astrophys. J.}\ }\textbf {\bibinfo {volume} {833}},\ \bibinfo {pages} {61}
  (\bibinfo {year} {2016})},\ \Eprint {http://arxiv.org/abs/1609.01588}
  {arXiv:1609.01588 [astro-ph.CO]} \BibitemShut {NoStop}%
\bibitem [{\citenamefont {Helfer}\ \emph {et~al.}(2017)\citenamefont {Helfer},
  \citenamefont {Marsh}, \citenamefont {Clough}, \citenamefont {Fairbairn},
  \citenamefont {Lim},\ and\ \citenamefont {Becerril}}]{Helfer:2016ljl}%
  \BibitemOpen
  \bibfield  {author} {\bibinfo {author} {\bibfnamefont {T.}~\bibnamefont
  {Helfer}}, \bibinfo {author} {\bibfnamefont {D.~J.~E.}\ \bibnamefont
  {Marsh}}, \bibinfo {author} {\bibfnamefont {K.}~\bibnamefont {Clough}},
  \bibinfo {author} {\bibfnamefont {M.}~\bibnamefont {Fairbairn}}, \bibinfo
  {author} {\bibfnamefont {E.~A.}\ \bibnamefont {Lim}}, \ and\ \bibinfo
  {author} {\bibfnamefont {R.}~\bibnamefont {Becerril}},\ }\href {\doibase
  10.1088/1475-7516/2017/03/055} {\bibfield  {journal} {\bibinfo  {journal}
  {JCAP}\ }\textbf {\bibinfo {volume} {03}},\ \bibinfo {pages} {055} (\bibinfo
  {year} {2017})},\ \Eprint {http://arxiv.org/abs/1609.04724} {arXiv:1609.04724
  [astro-ph.CO]} \BibitemShut {NoStop}%
\bibitem [{\citenamefont {Widdicombe}\ \emph {et~al.}(2018)\citenamefont
  {Widdicombe}, \citenamefont {Helfer}, \citenamefont {Marsh},\ and\
  \citenamefont {Lim}}]{Widdicombe:2018oeo}%
  \BibitemOpen
  \bibfield  {author} {\bibinfo {author} {\bibfnamefont {J.~Y.}\ \bibnamefont
  {Widdicombe}}, \bibinfo {author} {\bibfnamefont {T.}~\bibnamefont {Helfer}},
  \bibinfo {author} {\bibfnamefont {D.~J.~E.}\ \bibnamefont {Marsh}}, \ and\
  \bibinfo {author} {\bibfnamefont {E.~A.}\ \bibnamefont {Lim}},\ }\href
  {\doibase 10.1088/1475-7516/2018/10/005} {\bibfield  {journal} {\bibinfo
  {journal} {JCAP}\ }\textbf {\bibinfo {volume} {10}},\ \bibinfo {pages} {005}
  (\bibinfo {year} {2018})},\ \Eprint {http://arxiv.org/abs/1806.09367}
  {arXiv:1806.09367 [astro-ph.CO]} \BibitemShut {NoStop}%
\bibitem [{\citenamefont {Muia}\ \emph {et~al.}(2019)\citenamefont {Muia},
  \citenamefont {Cicoli}, \citenamefont {Clough}, \citenamefont {Pedro},
  \citenamefont {Quevedo},\ and\ \citenamefont {Vacca}}]{Muia:2019coe}%
  \BibitemOpen
  \bibfield  {author} {\bibinfo {author} {\bibfnamefont {F.}~\bibnamefont
  {Muia}}, \bibinfo {author} {\bibfnamefont {M.}~\bibnamefont {Cicoli}},
  \bibinfo {author} {\bibfnamefont {K.}~\bibnamefont {Clough}}, \bibinfo
  {author} {\bibfnamefont {F.}~\bibnamefont {Pedro}}, \bibinfo {author}
  {\bibfnamefont {F.}~\bibnamefont {Quevedo}}, \ and\ \bibinfo {author}
  {\bibfnamefont {G.~P.}\ \bibnamefont {Vacca}},\ }\href {\doibase
  10.1088/1475-7516/2019/07/044} {\bibfield  {journal} {\bibinfo  {journal}
  {JCAP}\ }\textbf {\bibinfo {volume} {07}},\ \bibinfo {pages} {044} (\bibinfo
  {year} {2019})},\ \Eprint {http://arxiv.org/abs/1906.09346} {arXiv:1906.09346
  [gr-qc]} \BibitemShut {NoStop}%
\bibitem [{\citenamefont {Nazari}\ \emph {et~al.}(2021)\citenamefont {Nazari},
  \citenamefont {Cicoli}, \citenamefont {Clough},\ and\ \citenamefont
  {Muia}}]{Nazari:2020fmk}%
  \BibitemOpen
  \bibfield  {author} {\bibinfo {author} {\bibfnamefont {Z.}~\bibnamefont
  {Nazari}}, \bibinfo {author} {\bibfnamefont {M.}~\bibnamefont {Cicoli}},
  \bibinfo {author} {\bibfnamefont {K.}~\bibnamefont {Clough}}, \ and\ \bibinfo
  {author} {\bibfnamefont {F.}~\bibnamefont {Muia}},\ }\href {\doibase
  10.1088/1475-7516/2021/05/027} {\bibfield  {journal} {\bibinfo  {journal}
  {JCAP}\ }\textbf {\bibinfo {volume} {05}},\ \bibinfo {pages} {027} (\bibinfo
  {year} {2021})},\ \Eprint {http://arxiv.org/abs/2010.05933} {arXiv:2010.05933
  [gr-qc]} \BibitemShut {NoStop}%
\bibitem [{\citenamefont {Eggemeier}\ \emph
  {et~al.}(2021{\natexlab{a}})\citenamefont {Eggemeier}, \citenamefont
  {Niemeyer},\ and\ \citenamefont {Easther}}]{Eggemeier:2020zeg}%
  \BibitemOpen
  \bibfield  {author} {\bibinfo {author} {\bibfnamefont {B.}~\bibnamefont
  {Eggemeier}}, \bibinfo {author} {\bibfnamefont {J.~C.}\ \bibnamefont
  {Niemeyer}}, \ and\ \bibinfo {author} {\bibfnamefont {R.}~\bibnamefont
  {Easther}},\ }\href {\doibase 10.1103/PhysRevD.103.063525} {\bibfield
  {journal} {\bibinfo  {journal} {Phys. Rev. D}\ }\textbf {\bibinfo {volume}
  {103}},\ \bibinfo {pages} {063525} (\bibinfo {year} {2021}{\natexlab{a}})},\
  \Eprint {http://arxiv.org/abs/2011.13333} {arXiv:2011.13333 [astro-ph.CO]}
  \BibitemShut {NoStop}%
\bibitem [{\citenamefont {Eggemeier}\ \emph
  {et~al.}(2021{\natexlab{b}})\citenamefont {Eggemeier}, \citenamefont
  {Schwabe}, \citenamefont {Niemeyer},\ and\ \citenamefont
  {Easther}}]{Eggemeier:2021smj}%
  \BibitemOpen
  \bibfield  {author} {\bibinfo {author} {\bibfnamefont {B.}~\bibnamefont
  {Eggemeier}}, \bibinfo {author} {\bibfnamefont {B.}~\bibnamefont {Schwabe}},
  \bibinfo {author} {\bibfnamefont {J.~C.}\ \bibnamefont {Niemeyer}}, \ and\
  \bibinfo {author} {\bibfnamefont {R.}~\bibnamefont {Easther}},\ }\href@noop
  {} {\  (\bibinfo {year} {2021}{\natexlab{b}})},\ \Eprint
  {http://arxiv.org/abs/2110.15109} {arXiv:2110.15109 [astro-ph.CO]}
  \BibitemShut {NoStop}%
\bibitem [{\citenamefont {Aghanim}\ \emph {et~al.}(2020)\citenamefont {Aghanim}
  \emph {et~al.}}]{Planck:2018vyg}%
  \BibitemOpen
  \bibfield  {author} {\bibinfo {author} {\bibfnamefont {N.}~\bibnamefont
  {Aghanim}} \emph {et~al.} (\bibinfo {collaboration} {Planck}),\ }\href
  {\doibase 10.1051/0004-6361/201833910} {\bibfield  {journal} {\bibinfo
  {journal} {Astron. Astrophys.}\ }\textbf {\bibinfo {volume} {641}},\ \bibinfo
  {pages} {A6} (\bibinfo {year} {2020})},\ \bibinfo {note} {[Erratum:
  Astron.Astrophys. 652, C4 (2021)]},\ \Eprint
  {http://arxiv.org/abs/1807.06209} {arXiv:1807.06209 [astro-ph.CO]}
  \BibitemShut {NoStop}%
\bibitem [{\citenamefont {Carr}\ and\ \citenamefont
  {Kuhnel}(2021)}]{Carr:2021bzv}%
  \BibitemOpen
  \bibfield  {author} {\bibinfo {author} {\bibfnamefont {B.}~\bibnamefont
  {Carr}}\ and\ \bibinfo {author} {\bibfnamefont {F.}~\bibnamefont {Kuhnel}},\
  }in\ \href@noop {} {\emph {\bibinfo {booktitle} {{Les Houches summer school
  on Dark Matter}}}}\ (\bibinfo {year} {2021})\ \Eprint
  {http://arxiv.org/abs/2110.02821} {arXiv:2110.02821 [astro-ph.CO]}
  \BibitemShut {NoStop}%
\bibitem [{\citenamefont {Carr}\ \emph {et~al.}(2021)\citenamefont {Carr},
  \citenamefont {Kohri}, \citenamefont {Sendouda},\ and\ \citenamefont
  {Yokoyama}}]{Carr:2020gox}%
  \BibitemOpen
  \bibfield  {author} {\bibinfo {author} {\bibfnamefont {B.}~\bibnamefont
  {Carr}}, \bibinfo {author} {\bibfnamefont {K.}~\bibnamefont {Kohri}},
  \bibinfo {author} {\bibfnamefont {Y.}~\bibnamefont {Sendouda}}, \ and\
  \bibinfo {author} {\bibfnamefont {J.}~\bibnamefont {Yokoyama}},\ }\href
  {\doibase 10.1088/1361-6633/ac1e31} {\bibfield  {journal} {\bibinfo
  {journal} {Rept. Prog. Phys.}\ }\textbf {\bibinfo {volume} {84}},\ \bibinfo
  {pages} {116902} (\bibinfo {year} {2021})},\ \Eprint
  {http://arxiv.org/abs/2002.12778} {arXiv:2002.12778 [astro-ph.CO]}
  \BibitemShut {NoStop}%
\bibitem [{\citenamefont {Calz\`a}\ \emph {et~al.}(2021)\citenamefont
  {Calz\`a}, \citenamefont {March-Russell},\ and\ \citenamefont
  {Rosa}}]{Calza:2021czr}%
  \BibitemOpen
  \bibfield  {author} {\bibinfo {author} {\bibfnamefont {M.}~\bibnamefont
  {Calz\`a}}, \bibinfo {author} {\bibfnamefont {J.}~\bibnamefont
  {March-Russell}}, \ and\ \bibinfo {author} {\bibfnamefont {J.~a.~G.}\
  \bibnamefont {Rosa}},\ }\href@noop {} {\  (\bibinfo {year} {2021})},\ \Eprint
  {http://arxiv.org/abs/2110.13602} {arXiv:2110.13602 [astro-ph.CO]}
  \BibitemShut {NoStop}%
\bibitem [{\citenamefont {Hooper}\ and\ \citenamefont
  {Krnjaic}(2021)}]{Hooper:2020otu}%
  \BibitemOpen
  \bibfield  {author} {\bibinfo {author} {\bibfnamefont {D.}~\bibnamefont
  {Hooper}}\ and\ \bibinfo {author} {\bibfnamefont {G.}~\bibnamefont
  {Krnjaic}},\ }\href {\doibase 10.1103/PhysRevD.103.043504} {\bibfield
  {journal} {\bibinfo  {journal} {Phys. Rev. D}\ }\textbf {\bibinfo {volume}
  {103}},\ \bibinfo {pages} {043504} (\bibinfo {year} {2021})},\ \Eprint
  {http://arxiv.org/abs/2010.01134} {arXiv:2010.01134 [hep-ph]} \BibitemShut
  {NoStop}%
\bibitem [{\citenamefont {Lennon}\ \emph {et~al.}(2018)\citenamefont {Lennon},
  \citenamefont {March-Russell}, \citenamefont {Petrossian-Byrne},\ and\
  \citenamefont {Tillim}}]{Lennon:2017tqq}%
  \BibitemOpen
  \bibfield  {author} {\bibinfo {author} {\bibfnamefont {O.}~\bibnamefont
  {Lennon}}, \bibinfo {author} {\bibfnamefont {J.}~\bibnamefont
  {March-Russell}}, \bibinfo {author} {\bibfnamefont {R.}~\bibnamefont
  {Petrossian-Byrne}}, \ and\ \bibinfo {author} {\bibfnamefont
  {H.}~\bibnamefont {Tillim}},\ }\href {\doibase 10.1088/1475-7516/2018/04/009}
  {\bibfield  {journal} {\bibinfo  {journal} {JCAP}\ }\textbf {\bibinfo
  {volume} {04}},\ \bibinfo {pages} {009} (\bibinfo {year} {2018})},\ \Eprint
  {http://arxiv.org/abs/1712.07664} {arXiv:1712.07664 [hep-ph]} \BibitemShut
  {NoStop}%
\bibitem [{\citenamefont {Baldes}\ \emph {et~al.}(2020)\citenamefont {Baldes},
  \citenamefont {Decant}, \citenamefont {Hooper},\ and\ \citenamefont
  {Lopez-Honorez}}]{Baldes:2020nuv}%
  \BibitemOpen
  \bibfield  {author} {\bibinfo {author} {\bibfnamefont {I.}~\bibnamefont
  {Baldes}}, \bibinfo {author} {\bibfnamefont {Q.}~\bibnamefont {Decant}},
  \bibinfo {author} {\bibfnamefont {D.~C.}\ \bibnamefont {Hooper}}, \ and\
  \bibinfo {author} {\bibfnamefont {L.}~\bibnamefont {Lopez-Honorez}},\ }\href
  {\doibase 10.1088/1475-7516/2020/08/045} {\bibfield  {journal} {\bibinfo
  {journal} {JCAP}\ }\textbf {\bibinfo {volume} {08}},\ \bibinfo {pages} {045}
  (\bibinfo {year} {2020})},\ \Eprint {http://arxiv.org/abs/2004.14773}
  {arXiv:2004.14773 [astro-ph.CO]} \BibitemShut {NoStop}%
\bibitem [{\citenamefont {Anantua}\ \emph {et~al.}(2009)\citenamefont
  {Anantua}, \citenamefont {Easther},\ and\ \citenamefont
  {Giblin}}]{Anantua:2008am}%
  \BibitemOpen
  \bibfield  {author} {\bibinfo {author} {\bibfnamefont {R.}~\bibnamefont
  {Anantua}}, \bibinfo {author} {\bibfnamefont {R.}~\bibnamefont {Easther}}, \
  and\ \bibinfo {author} {\bibfnamefont {J.~T.}\ \bibnamefont {Giblin}},\
  }\href {\doibase 10.1103/PhysRevLett.103.111303} {\bibfield  {journal}
  {\bibinfo  {journal} {Phys. Rev. Lett.}\ }\textbf {\bibinfo {volume} {103}},\
  \bibinfo {pages} {111303} (\bibinfo {year} {2009})},\ \Eprint
  {http://arxiv.org/abs/0812.0825} {arXiv:0812.0825 [astro-ph]} \BibitemShut
  {NoStop}%
\bibitem [{\citenamefont {Dolgov}\ and\ \citenamefont
  {Ejlli}(2011)}]{Dolgov:2011cq}%
  \BibitemOpen
  \bibfield  {author} {\bibinfo {author} {\bibfnamefont {A.~D.}\ \bibnamefont
  {Dolgov}}\ and\ \bibinfo {author} {\bibfnamefont {D.}~\bibnamefont {Ejlli}},\
  }\href {\doibase 10.1103/PhysRevD.84.024028} {\bibfield  {journal} {\bibinfo
  {journal} {Phys. Rev. D}\ }\textbf {\bibinfo {volume} {84}},\ \bibinfo
  {pages} {024028} (\bibinfo {year} {2011})},\ \Eprint
  {http://arxiv.org/abs/1105.2303} {arXiv:1105.2303 [astro-ph.CO]} \BibitemShut
  {NoStop}%
\bibitem [{\citenamefont {Zagorac}\ \emph {et~al.}(2019)\citenamefont
  {Zagorac}, \citenamefont {Easther},\ and\ \citenamefont
  {Padmanabhan}}]{Zagorac:2019ekv}%
  \BibitemOpen
  \bibfield  {author} {\bibinfo {author} {\bibfnamefont {J.~L.}\ \bibnamefont
  {Zagorac}}, \bibinfo {author} {\bibfnamefont {R.}~\bibnamefont {Easther}}, \
  and\ \bibinfo {author} {\bibfnamefont {N.}~\bibnamefont {Padmanabhan}},\
  }\href {\doibase 10.1088/1475-7516/2019/06/052} {\bibfield  {journal}
  {\bibinfo  {journal} {JCAP}\ }\textbf {\bibinfo {volume} {06}},\ \bibinfo
  {pages} {052} (\bibinfo {year} {2019})},\ \Eprint
  {http://arxiv.org/abs/1903.05053} {arXiv:1903.05053 [astro-ph.CO]}
  \BibitemShut {NoStop}%
\bibitem [{\citenamefont {Blinov}\ \emph {et~al.}(2021)\citenamefont {Blinov},
  \citenamefont {Dolan}, \citenamefont {Draper},\ and\ \citenamefont
  {Shelton}}]{Blinov:2021axd}%
  \BibitemOpen
  \bibfield  {author} {\bibinfo {author} {\bibfnamefont {N.}~\bibnamefont
  {Blinov}}, \bibinfo {author} {\bibfnamefont {M.~J.}\ \bibnamefont {Dolan}},
  \bibinfo {author} {\bibfnamefont {P.}~\bibnamefont {Draper}}, \ and\ \bibinfo
  {author} {\bibfnamefont {J.}~\bibnamefont {Shelton}},\ }\href {\doibase
  10.1103/PhysRevD.103.103514} {\bibfield  {journal} {\bibinfo  {journal}
  {Phys. Rev. D}\ }\textbf {\bibinfo {volume} {103}},\ \bibinfo {pages}
  {103514} (\bibinfo {year} {2021})},\ \Eprint
  {http://arxiv.org/abs/2102.05070} {arXiv:2102.05070 [astro-ph.CO]}
  \BibitemShut {NoStop}%
\bibitem [{\citenamefont {Barenboim}\ \emph {et~al.}(2021)\citenamefont
  {Barenboim}, \citenamefont {Blinov},\ and\ \citenamefont
  {Stebbins}}]{Barenboim:2021swl}%
  \BibitemOpen
  \bibfield  {author} {\bibinfo {author} {\bibfnamefont {G.}~\bibnamefont
  {Barenboim}}, \bibinfo {author} {\bibfnamefont {N.}~\bibnamefont {Blinov}}, \
  and\ \bibinfo {author} {\bibfnamefont {A.}~\bibnamefont {Stebbins}},\ }\href
  {\doibase 10.1088/1475-7516/2021/12/026} {\bibfield  {journal} {\bibinfo
  {journal} {JCAP}\ }\textbf {\bibinfo {volume} {12}},\ \bibinfo {pages} {026}
  (\bibinfo {year} {2021})},\ \Eprint {http://arxiv.org/abs/2107.10293}
  {arXiv:2107.10293 [astro-ph.CO]} \BibitemShut {NoStop}%
\bibitem [{\citenamefont {Blanco}\ \emph {et~al.}(2019)\citenamefont {Blanco},
  \citenamefont {Delos}, \citenamefont {Erickcek},\ and\ \citenamefont
  {Hooper}}]{Blanco:2019eij}%
  \BibitemOpen
  \bibfield  {author} {\bibinfo {author} {\bibfnamefont {C.}~\bibnamefont
  {Blanco}}, \bibinfo {author} {\bibfnamefont {M.~S.}\ \bibnamefont {Delos}},
  \bibinfo {author} {\bibfnamefont {A.~L.}\ \bibnamefont {Erickcek}}, \ and\
  \bibinfo {author} {\bibfnamefont {D.}~\bibnamefont {Hooper}},\ }\href
  {\doibase 10.1103/PhysRevD.100.103010} {\bibfield  {journal} {\bibinfo
  {journal} {Phys. Rev. D}\ }\textbf {\bibinfo {volume} {100}},\ \bibinfo
  {pages} {103010} (\bibinfo {year} {2019})},\ \Eprint
  {http://arxiv.org/abs/1906.00010} {arXiv:1906.00010 [astro-ph.CO]}
  \BibitemShut {NoStop}%
\bibitem [{\citenamefont {Flores}\ and\ \citenamefont
  {Kusenko}(2021)}]{Flores:2020drq}%
  \BibitemOpen
  \bibfield  {author} {\bibinfo {author} {\bibfnamefont {M.~M.}\ \bibnamefont
  {Flores}}\ and\ \bibinfo {author} {\bibfnamefont {A.}~\bibnamefont
  {Kusenko}},\ }\href {\doibase 10.1103/PhysRevLett.126.041101} {\bibfield
  {journal} {\bibinfo  {journal} {Phys. Rev. Lett.}\ }\textbf {\bibinfo
  {volume} {126}},\ \bibinfo {pages} {041101} (\bibinfo {year} {2021})},\
  \Eprint {http://arxiv.org/abs/2008.12456} {arXiv:2008.12456 [astro-ph.CO]}
  \BibitemShut {NoStop}%
\bibitem [{\citenamefont {Baumann}\ \emph {et~al.}(2007)\citenamefont
  {Baumann}, \citenamefont {Steinhardt}, \citenamefont {Takahashi},\ and\
  \citenamefont {Ichiki}}]{Baumann:2007zm}%
  \BibitemOpen
  \bibfield  {author} {\bibinfo {author} {\bibfnamefont {D.}~\bibnamefont
  {Baumann}}, \bibinfo {author} {\bibfnamefont {P.~J.}\ \bibnamefont
  {Steinhardt}}, \bibinfo {author} {\bibfnamefont {K.}~\bibnamefont
  {Takahashi}}, \ and\ \bibinfo {author} {\bibfnamefont {K.}~\bibnamefont
  {Ichiki}},\ }\href {\doibase 10.1103/PhysRevD.76.084019} {\bibfield
  {journal} {\bibinfo  {journal} {Phys. Rev. D}\ }\textbf {\bibinfo {volume}
  {76}},\ \bibinfo {pages} {084019} (\bibinfo {year} {2007})},\ \Eprint
  {http://arxiv.org/abs/hep-th/0703290} {arXiv:hep-th/0703290} \BibitemShut
  {NoStop}%
\bibitem [{\citenamefont {Assadullahi}\ and\ \citenamefont
  {Wands}(2009)}]{Assadullahi:2009nf}%
  \BibitemOpen
  \bibfield  {author} {\bibinfo {author} {\bibfnamefont {H.}~\bibnamefont
  {Assadullahi}}\ and\ \bibinfo {author} {\bibfnamefont {D.}~\bibnamefont
  {Wands}},\ }\href {\doibase 10.1103/PhysRevD.79.083511} {\bibfield  {journal}
  {\bibinfo  {journal} {Phys. Rev. D}\ }\textbf {\bibinfo {volume} {79}},\
  \bibinfo {pages} {083511} (\bibinfo {year} {2009})},\ \Eprint
  {http://arxiv.org/abs/0901.0989} {arXiv:0901.0989 [astro-ph.CO]} \BibitemShut
  {NoStop}%
\bibitem [{\citenamefont {Espinosa}\ \emph {et~al.}(2018)\citenamefont
  {Espinosa}, \citenamefont {Racco},\ and\ \citenamefont
  {Riotto}}]{Espinosa:2018eve}%
  \BibitemOpen
  \bibfield  {author} {\bibinfo {author} {\bibfnamefont {J.~R.}\ \bibnamefont
  {Espinosa}}, \bibinfo {author} {\bibfnamefont {D.}~\bibnamefont {Racco}}, \
  and\ \bibinfo {author} {\bibfnamefont {A.}~\bibnamefont {Riotto}},\ }\href
  {\doibase 10.1088/1475-7516/2018/09/012} {\bibfield  {journal} {\bibinfo
  {journal} {JCAP}\ }\textbf {\bibinfo {volume} {09}},\ \bibinfo {pages} {012}
  (\bibinfo {year} {2018})},\ \Eprint {http://arxiv.org/abs/1804.07732}
  {arXiv:1804.07732 [hep-ph]} \BibitemShut {NoStop}%
\bibitem [{\citenamefont {Kohri}\ and\ \citenamefont
  {Terada}(2018)}]{Kohri:2018awv}%
  \BibitemOpen
  \bibfield  {author} {\bibinfo {author} {\bibfnamefont {K.}~\bibnamefont
  {Kohri}}\ and\ \bibinfo {author} {\bibfnamefont {T.}~\bibnamefont {Terada}},\
  }\href {\doibase 10.1103/PhysRevD.97.123532} {\bibfield  {journal} {\bibinfo
  {journal} {Phys. Rev. D}\ }\textbf {\bibinfo {volume} {97}},\ \bibinfo
  {pages} {123532} (\bibinfo {year} {2018})},\ \Eprint
  {http://arxiv.org/abs/1804.08577} {arXiv:1804.08577 [gr-qc]} \BibitemShut
  {NoStop}%
\bibitem [{\citenamefont {Inomata}\ \emph
  {et~al.}(2020{\natexlab{a}})\citenamefont {Inomata}, \citenamefont {Kohri},
  \citenamefont {Nakama},\ and\ \citenamefont {Terada}}]{Inomata:2020yqv}%
  \BibitemOpen
  \bibfield  {author} {\bibinfo {author} {\bibfnamefont {K.}~\bibnamefont
  {Inomata}}, \bibinfo {author} {\bibfnamefont {K.}~\bibnamefont {Kohri}},
  \bibinfo {author} {\bibfnamefont {T.}~\bibnamefont {Nakama}}, \ and\ \bibinfo
  {author} {\bibfnamefont {T.}~\bibnamefont {Terada}},\ }\href {\doibase
  10.1088/1742-6596/1468/1/012001} {\bibfield  {journal} {\bibinfo  {journal}
  {J. Phys. Conf. Ser.}\ }\textbf {\bibinfo {volume} {1468}},\ \bibinfo {pages}
  {012001} (\bibinfo {year} {2020}{\natexlab{a}})}\BibitemShut {NoStop}%
\bibitem [{\citenamefont {Inomata}\ \emph
  {et~al.}(2020{\natexlab{b}})\citenamefont {Inomata}, \citenamefont {Kohri},
  \citenamefont {Nakama},\ and\ \citenamefont {Terada}}]{Inomata:2020tkl}%
  \BibitemOpen
  \bibfield  {author} {\bibinfo {author} {\bibfnamefont {K.}~\bibnamefont
  {Inomata}}, \bibinfo {author} {\bibfnamefont {K.}~\bibnamefont {Kohri}},
  \bibinfo {author} {\bibfnamefont {T.}~\bibnamefont {Nakama}}, \ and\ \bibinfo
  {author} {\bibfnamefont {T.}~\bibnamefont {Terada}},\ }\href {\doibase
  10.1088/1742-6596/1468/1/012002} {\bibfield  {journal} {\bibinfo  {journal}
  {J. Phys. Conf. Ser.}\ }\textbf {\bibinfo {volume} {1468}},\ \bibinfo {pages}
  {012002} (\bibinfo {year} {2020}{\natexlab{b}})}\BibitemShut {NoStop}%
\bibitem [{\citenamefont {Dom\`enech}\ \emph
  {et~al.}(2021{\natexlab{a}})\citenamefont {Dom\`enech}, \citenamefont {Lin},\
  and\ \citenamefont {Sasaki}}]{Domenech:2020ssp}%
  \BibitemOpen
  \bibfield  {author} {\bibinfo {author} {\bibfnamefont {G.}~\bibnamefont
  {Dom\`enech}}, \bibinfo {author} {\bibfnamefont {C.}~\bibnamefont {Lin}}, \
  and\ \bibinfo {author} {\bibfnamefont {M.}~\bibnamefont {Sasaki}},\ }\href
  {\doibase 10.1088/1475-7516/2021/11/E01} {\bibfield  {journal} {\bibinfo
  {journal} {JCAP}\ }\textbf {\bibinfo {volume} {11}},\ \bibinfo {pages} {E01}
  (\bibinfo {year} {2021}{\natexlab{a}})},\ \Eprint
  {http://arxiv.org/abs/2012.08151} {arXiv:2012.08151 [gr-qc]} \BibitemShut
  {NoStop}%
\bibitem [{\citenamefont {Dom\`enech}\ \emph
  {et~al.}(2021{\natexlab{b}})\citenamefont {Dom\`enech}, \citenamefont
  {Takhistov},\ and\ \citenamefont {Sasaki}}]{Domenech:2021wkk}%
  \BibitemOpen
  \bibfield  {author} {\bibinfo {author} {\bibfnamefont {G.}~\bibnamefont
  {Dom\`enech}}, \bibinfo {author} {\bibfnamefont {V.}~\bibnamefont
  {Takhistov}}, \ and\ \bibinfo {author} {\bibfnamefont {M.}~\bibnamefont
  {Sasaki}},\ }\href {\doibase 10.1016/j.physletb.2021.136722} {\bibfield
  {journal} {\bibinfo  {journal} {Phys. Lett. B}\ }\textbf {\bibinfo {volume}
  {823}},\ \bibinfo {pages} {136722} (\bibinfo {year} {2021}{\natexlab{b}})},\
  \Eprint {http://arxiv.org/abs/2105.06816} {arXiv:2105.06816 [astro-ph.CO]}
  \BibitemShut {NoStop}%
\bibitem [{\citenamefont {Dom\`enech}(2021)}]{Domenech:2021ztg}%
  \BibitemOpen
  \bibfield  {author} {\bibinfo {author} {\bibfnamefont {G.}~\bibnamefont
  {Dom\`enech}},\ }\href {\doibase 10.3390/universe7110398} {\bibfield
  {journal} {\bibinfo  {journal} {Universe}\ }\textbf {\bibinfo {volume} {7}},\
  \bibinfo {pages} {398} (\bibinfo {year} {2021})},\ \Eprint
  {http://arxiv.org/abs/2109.01398} {arXiv:2109.01398 [gr-qc]} \BibitemShut
  {NoStop}%
\bibitem [{\citenamefont {Giblin}\ and\ \citenamefont
  {Thrane}(2014)}]{Giblin:2014gra}%
  \BibitemOpen
  \bibfield  {author} {\bibinfo {author} {\bibfnamefont {J.~T.}\ \bibnamefont
  {Giblin}}\ and\ \bibinfo {author} {\bibfnamefont {E.}~\bibnamefont
  {Thrane}},\ }\href {\doibase 10.1103/PhysRevD.90.107502} {\bibfield
  {journal} {\bibinfo  {journal} {Phys. Rev. D}\ }\textbf {\bibinfo {volume}
  {90}},\ \bibinfo {pages} {107502} (\bibinfo {year} {2014})},\ \Eprint
  {http://arxiv.org/abs/1410.4779} {arXiv:1410.4779 [gr-qc]} \BibitemShut
  {NoStop}%
\bibitem [{\citenamefont {Chatrchyan}\ and\ \citenamefont
  {Jaeckel}(2021)}]{Chatrchyan:2020pzh}%
  \BibitemOpen
  \bibfield  {author} {\bibinfo {author} {\bibfnamefont {A.}~\bibnamefont
  {Chatrchyan}}\ and\ \bibinfo {author} {\bibfnamefont {J.}~\bibnamefont
  {Jaeckel}},\ }\href {\doibase 10.1088/1475-7516/2021/02/003} {\bibfield
  {journal} {\bibinfo  {journal} {JCAP}\ }\textbf {\bibinfo {volume} {02}},\
  \bibinfo {pages} {003} (\bibinfo {year} {2021})},\ \Eprint
  {http://arxiv.org/abs/2004.07844} {arXiv:2004.07844 [hep-ph]} \BibitemShut
  {NoStop}%
\bibitem [{\citenamefont {et~al.}(2017)}]{amaroseoane2017laser}%
  \BibitemOpen
  \bibfield  {author} {\bibinfo {author} {\bibfnamefont {P.~A.-S.}\
  \bibnamefont {et~al.}},\ }\href@noop {} {\enquote {\bibinfo {title} {Laser
  interferometer space antenna},}\ } (\bibinfo {year} {2017}),\ \Eprint
  {http://arxiv.org/abs/1702.00786} {arXiv:1702.00786 [astro-ph.IM]}
  \BibitemShut {NoStop}%
\bibitem [{\citenamefont {Barausse}\ \emph {et~al.}(2020)\citenamefont
  {Barausse} \emph {et~al.}}]{Barausse:2020rsu}%
  \BibitemOpen
  \bibfield  {author} {\bibinfo {author} {\bibfnamefont {E.}~\bibnamefont
  {Barausse}} \emph {et~al.},\ }\href {\doibase 10.1007/s10714-020-02691-1}
  {\bibfield  {journal} {\bibinfo  {journal} {Gen. Rel. Grav.}\ }\textbf
  {\bibinfo {volume} {52}},\ \bibinfo {pages} {81} (\bibinfo {year} {2020})},\
  \Eprint {http://arxiv.org/abs/2001.09793} {arXiv:2001.09793 [gr-qc]}
  \BibitemShut {NoStop}%
\bibitem [{\citenamefont {Seto}\ \emph {et~al.}(2001)\citenamefont {Seto},
  \citenamefont {Kawamura},\ and\ \citenamefont {Nakamura}}]{Seto:2001qf}%
  \BibitemOpen
  \bibfield  {author} {\bibinfo {author} {\bibfnamefont {N.}~\bibnamefont
  {Seto}}, \bibinfo {author} {\bibfnamefont {S.}~\bibnamefont {Kawamura}}, \
  and\ \bibinfo {author} {\bibfnamefont {T.}~\bibnamefont {Nakamura}},\ }\href
  {\doibase 10.1103/PhysRevLett.87.221103} {\bibfield  {journal} {\bibinfo
  {journal} {Phys. Rev. Lett.}\ }\textbf {\bibinfo {volume} {87}},\ \bibinfo
  {pages} {221103} (\bibinfo {year} {2001})},\ \Eprint
  {http://arxiv.org/abs/astro-ph/0108011} {arXiv:astro-ph/0108011} \BibitemShut
  {NoStop}%
\bibitem [{\citenamefont {Yagi}\ and\ \citenamefont
  {Seto}(2011)}]{Yagi:2011wg}%
  \BibitemOpen
  \bibfield  {author} {\bibinfo {author} {\bibfnamefont {K.}~\bibnamefont
  {Yagi}}\ and\ \bibinfo {author} {\bibfnamefont {N.}~\bibnamefont {Seto}},\
  }\href {\doibase 10.1103/PhysRevD.83.044011} {\bibfield  {journal} {\bibinfo
  {journal} {Phys. Rev. D}\ }\textbf {\bibinfo {volume} {83}},\ \bibinfo
  {pages} {044011} (\bibinfo {year} {2011})},\ \bibinfo {note} {[Erratum:
  Phys.Rev.D 95, 109901 (2017)]},\ \Eprint {http://arxiv.org/abs/1101.3940}
  {arXiv:1101.3940 [astro-ph.CO]} \BibitemShut {NoStop}%
\bibitem [{\citenamefont {Abbott}\ \emph {et~al.}(2021)\citenamefont {Abbott}
  \emph {et~al.}}]{KAGRA:2021kbb}%
  \BibitemOpen
  \bibfield  {author} {\bibinfo {author} {\bibfnamefont {R.}~\bibnamefont
  {Abbott}} \emph {et~al.} (\bibinfo {collaboration} {KAGRA, Virgo, LIGO
  Scientific}),\ }\href {\doibase 10.1103/PhysRevD.104.022004} {\bibfield
  {journal} {\bibinfo  {journal} {Phys. Rev. D}\ }\textbf {\bibinfo {volume}
  {104}},\ \bibinfo {pages} {022004} (\bibinfo {year} {2021})},\ \Eprint
  {http://arxiv.org/abs/2101.12130} {arXiv:2101.12130 [gr-qc]} \BibitemShut
  {NoStop}%
\bibitem [{\citenamefont {Maggiore}\ \emph {et~al.}(2020)\citenamefont
  {Maggiore} \emph {et~al.}}]{Maggiore:2019uih}%
  \BibitemOpen
  \bibfield  {author} {\bibinfo {author} {\bibfnamefont {M.}~\bibnamefont
  {Maggiore}} \emph {et~al.},\ }\href {\doibase 10.1088/1475-7516/2020/03/050}
  {\bibfield  {journal} {\bibinfo  {journal} {JCAP}\ }\textbf {\bibinfo
  {volume} {03}},\ \bibinfo {pages} {050} (\bibinfo {year} {2020})},\ \Eprint
  {http://arxiv.org/abs/1912.02622} {arXiv:1912.02622 [astro-ph.CO]}
  \BibitemShut {NoStop}%
\bibitem [{\citenamefont {Evans}\ \emph {et~al.}(2021)\citenamefont {Evans}
  \emph {et~al.}}]{Evans:2021gyd}%
  \BibitemOpen
  \bibfield  {author} {\bibinfo {author} {\bibfnamefont {M.}~\bibnamefont
  {Evans}} \emph {et~al.},\ }\href@noop {} {\  (\bibinfo {year} {2021})},\
  \Eprint {http://arxiv.org/abs/2109.09882} {arXiv:2109.09882 [astro-ph.IM]}
  \BibitemShut {NoStop}%
\bibitem [{\citenamefont {Aggarwal}\ \emph {et~al.}(2021)\citenamefont
  {Aggarwal} \emph {et~al.}}]{Aggarwal:2020olq}%
  \BibitemOpen
  \bibfield  {author} {\bibinfo {author} {\bibfnamefont {N.}~\bibnamefont
  {Aggarwal}} \emph {et~al.},\ }\href {\doibase 10.1007/s41114-021-00032-5}
  {\bibfield  {journal} {\bibinfo  {journal} {Living Rev. Rel.}\ }\textbf
  {\bibinfo {volume} {24}},\ \bibinfo {pages} {4} (\bibinfo {year} {2021})},\
  \Eprint {http://arxiv.org/abs/2011.12414} {arXiv:2011.12414 [gr-qc]}
  \BibitemShut {NoStop}%
\bibitem [{\citenamefont {Delos}(2020)}]{Delos:2020mtj}%
  \BibitemOpen
  \bibfield  {author} {\bibinfo {author} {\bibfnamefont {M.~S.}\ \bibnamefont
  {Delos}},\ }\emph {\bibinfo {title} {{Probing the Early Universe Using Dark
  Matter Minihalos}}},\ \href {\doibase 10.17615/0j00-h373} {Ph.D. thesis},\
  \bibinfo  {school} {UNC, Chapel Hill} (\bibinfo {year} {2020})\BibitemShut
  {NoStop}%
\bibitem [{\citenamefont {Delos}(2019)}]{Delos:2019tsl}%
  \BibitemOpen
  \bibfield  {author} {\bibinfo {author} {\bibfnamefont {M.~S.}\ \bibnamefont
  {Delos}},\ }\href {\doibase 10.1103/PhysRevD.100.083529} {\bibfield
  {journal} {\bibinfo  {journal} {Phys. Rev. D}\ }\textbf {\bibinfo {volume}
  {100}},\ \bibinfo {pages} {083529} (\bibinfo {year} {2019})},\ \Eprint
  {http://arxiv.org/abs/1907.13133} {arXiv:1907.13133 [astro-ph.CO]}
  \BibitemShut {NoStop}%
\bibitem [{\citenamefont {Delos}\ \emph {et~al.}(2018)\citenamefont {Delos},
  \citenamefont {Erickcek}, \citenamefont {Bailey},\ and\ \citenamefont
  {Alvarez}}]{Delos:2018ueo}%
  \BibitemOpen
  \bibfield  {author} {\bibinfo {author} {\bibfnamefont {M.~S.}\ \bibnamefont
  {Delos}}, \bibinfo {author} {\bibfnamefont {A.~L.}\ \bibnamefont {Erickcek}},
  \bibinfo {author} {\bibfnamefont {A.~P.}\ \bibnamefont {Bailey}}, \ and\
  \bibinfo {author} {\bibfnamefont {M.~A.}\ \bibnamefont {Alvarez}},\ }\href
  {\doibase 10.1103/PhysRevD.98.063527} {\bibfield  {journal} {\bibinfo
  {journal} {Phys. Rev. D}\ }\textbf {\bibinfo {volume} {98}},\ \bibinfo
  {pages} {063527} (\bibinfo {year} {2018})},\ \Eprint
  {http://arxiv.org/abs/1806.07389} {arXiv:1806.07389 [astro-ph.CO]}
  \BibitemShut {NoStop}%
\bibitem [{\citenamefont {Thrane}\ and\ \citenamefont
  {Romano}(2013)}]{Thrane:2013oya}%
  \BibitemOpen
  \bibfield  {author} {\bibinfo {author} {\bibfnamefont {E.}~\bibnamefont
  {Thrane}}\ and\ \bibinfo {author} {\bibfnamefont {J.~D.}\ \bibnamefont
  {Romano}},\ }\href {\doibase 10.1103/PhysRevD.88.124032} {\bibfield
  {journal} {\bibinfo  {journal} {Phys. Rev. D}\ }\textbf {\bibinfo {volume}
  {88}},\ \bibinfo {pages} {124032} (\bibinfo {year} {2013})},\ \Eprint
  {http://arxiv.org/abs/1310.5300} {arXiv:1310.5300 [astro-ph.IM]} \BibitemShut
  {NoStop}%
\bibitem [{\citenamefont {Mingarelli}\ \emph {et~al.}(2019)\citenamefont
  {Mingarelli}, \citenamefont {Taylor}, \citenamefont {Sathyaprakash},\ and\
  \citenamefont {Farr}}]{Mingarelli:2019mvk}%
  \BibitemOpen
  \bibfield  {author} {\bibinfo {author} {\bibfnamefont {C.~M.~F.}\
  \bibnamefont {Mingarelli}}, \bibinfo {author} {\bibfnamefont {S.~R.}\
  \bibnamefont {Taylor}}, \bibinfo {author} {\bibfnamefont {B.~S.}\
  \bibnamefont {Sathyaprakash}}, \ and\ \bibinfo {author} {\bibfnamefont
  {W.~M.}\ \bibnamefont {Farr}},\ }\href@noop {} {\  (\bibinfo {year}
  {2019})},\ \Eprint {http://arxiv.org/abs/1911.09745} {arXiv:1911.09745
  [gr-qc]} \BibitemShut {NoStop}%
\bibitem [{\citenamefont {Ejlli}\ \emph {et~al.}(2019)\citenamefont {Ejlli},
  \citenamefont {Ejlli}, \citenamefont {Cruise}, \citenamefont {Pisano},\ and\
  \citenamefont {Grote}}]{Ejlli:2019bqj}%
  \BibitemOpen
  \bibfield  {author} {\bibinfo {author} {\bibfnamefont {A.}~\bibnamefont
  {Ejlli}}, \bibinfo {author} {\bibfnamefont {D.}~\bibnamefont {Ejlli}},
  \bibinfo {author} {\bibfnamefont {A.~M.}\ \bibnamefont {Cruise}}, \bibinfo
  {author} {\bibfnamefont {G.}~\bibnamefont {Pisano}}, \ and\ \bibinfo {author}
  {\bibfnamefont {H.}~\bibnamefont {Grote}},\ }\href {\doibase
  10.1140/epjc/s10052-019-7542-5} {\bibfield  {journal} {\bibinfo  {journal}
  {Eur. Phys. J. C}\ }\textbf {\bibinfo {volume} {79}},\ \bibinfo {pages}
  {1032} (\bibinfo {year} {2019})},\ \Eprint {http://arxiv.org/abs/1908.00232}
  {arXiv:1908.00232 [gr-qc]} \BibitemShut {NoStop}%
\bibitem [{\citenamefont {B\"ahre}\ \emph {et~al.}(2013)\citenamefont {B\"ahre}
  \emph {et~al.}}]{Bahre:2013ywa}%
  \BibitemOpen
  \bibfield  {author} {\bibinfo {author} {\bibfnamefont {R.}~\bibnamefont
  {B\"ahre}} \emph {et~al.},\ }\href {\doibase 10.1088/1748-0221/8/09/T09001}
  {\bibfield  {journal} {\bibinfo  {journal} {JINST}\ }\textbf {\bibinfo
  {volume} {8}},\ \bibinfo {pages} {T09001} (\bibinfo {year} {2013})},\ \Eprint
  {http://arxiv.org/abs/1302.5647} {arXiv:1302.5647 [physics.ins-det]}
  \BibitemShut {NoStop}%
\bibitem [{\citenamefont {Graham}\ \emph {et~al.}(2015)\citenamefont {Graham},
  \citenamefont {Irastorza}, \citenamefont {Lamoreaux}, \citenamefont
  {Lindner},\ and\ \citenamefont {van
  Bibber}}]{doi:10.1146/annurev-nucl-102014-022120}%
  \BibitemOpen
  \bibfield  {author} {\bibinfo {author} {\bibfnamefont {P.~W.}\ \bibnamefont
  {Graham}}, \bibinfo {author} {\bibfnamefont {I.~G.}\ \bibnamefont
  {Irastorza}}, \bibinfo {author} {\bibfnamefont {S.~K.}\ \bibnamefont
  {Lamoreaux}}, \bibinfo {author} {\bibfnamefont {A.}~\bibnamefont {Lindner}},
  \ and\ \bibinfo {author} {\bibfnamefont {K.~A.}\ \bibnamefont {van Bibber}},\
  }\href {\doibase 10.1146/annurev-nucl-102014-022120} {\bibfield  {journal}
  {\bibinfo  {journal} {Annual Review of Nuclear and Particle Science}\
  }\textbf {\bibinfo {volume} {65}},\ \bibinfo {pages} {485} (\bibinfo {year}
  {2015})},\ \Eprint
  {http://arxiv.org/abs/https://doi.org/10.1146/annurev-nucl-102014-022120}
  {https://doi.org/10.1146/annurev-nucl-102014-022120} \BibitemShut {NoStop}%
\end{thebibliography}%
\end{document}